# IMPROVED FULLY AUTOMATED METHOD FOR THE DETERMINATION OF MEDIUM TO HIGHLY POLAR PESTICIDES IN SURFACE AND GROUNDWATER AND APPLICATION IN TWO DISTINCT AGRICULTURE-IMPACTED AREAS.

Maria Vittoria Barbieri[a], Luis Simón Monllor-Alcaraz[a], Cristina Postigo[a*], Miren López de Alda[a*]

[a]*Water, Environmental and Food Chemistry Unit (ENFOCHEM), Department of Environmental Chemistry, Institute of Environmental Assessment and Water Research (IDAEA-CSIC), Barcelona, Spain*

**Corresponding authors:*

Cristina Postigo (0000-0002-7344-7044)
cprqam@cid.csic.es
Miren López de Alda (0000-0002-9347-2765)
mlaqam@cid.csic.es
Institute of Environmental Assessment and Water Research (IDAEA-CSIC)
Department of Environmental Chemistry
C/ Jordi Girona 18-26, 08034 Barcelona, Spain.
Tel: +34-934-006-100, Fax: +34-932-045-904

**ABSTRACT**

Water is an essential resource for all living organisms. The continuous and increasing use of pesticides in agricultural and urban activities results in the pollution of water resources and represents an environmental risk. To control and reduce pesticide pollution, reliable multi-residue methods for the detection of these compounds in water are needed. In this context, the present work aimed at providing an analytical method for the simultaneous determination of trace levels of 51 target pesticides in water and applying it to the investigation of target pesticides in two agriculture-impacted areas of interest. The method developed, based on an isotopic dilution approach and on-line solid-phase extraction-liquid chromatography-tandem mass spectrometry, is fast, simple, and to a large extent automated, and allows the analysis of most of the target compounds in compliance with European regulations. Further application of the method to the analysis of selected water samples collected at the lowest stretches of the two largest river basins of Catalonia (NE Spain), Llobregat and Ter, revealed the presence of a wide suite of pesticides, and some of them at concentrations above the water quality standards (irgarol and dichlorvos) or the acceptable method detection limits (methiocarb, imidacloprid, and thiacloprid), in the Llobregat, and much cleaner waters in the Ter River basin. Risk assessment of the pesticide concentrations measured in the Llobregat indicated high risk due to the presence of irgarol, dichlorvos, methiocarb, azinphos ethyl, imidacloprid, and diflufenican (hazard quotient (HQ) values>10), and an only moderate potential risk in the Ter River associated to the occurrence of bentazone and irgarol (HQ>1).

## 1. Introduction

The extensive, and sometimes excessive, use of pesticides has led to significant undesired effects on the environmental and human health (Han et al., 2018; Islam et al., 2018; Mostafalou and Abdollahi, 2017). As a result, developed countries have withdrawn from the market the most toxic and persistent pesticides, such as the organochlorine ones, and have promoted the use of comparatively more polar substances, expected to degrade more rapidly and to be less toxic to non-target organisms (Sander et al., 2017). However, their use in large amounts has still resulted in their accumulation in the environment. Nowadays, over 500 active ingredients are available in the European market, with an estimated total turnover of around 400,000 tons (Eurostat, 2020). Spain is ranked as the first country in Europe with the largest pesticide consumption (according to pesticide sales (Eurostat, 2020), 72 M kg on average in the period 2011-2018).

Water plays an important role in the environmental fate of pesticides, because it transports these substances from agricultural to other areas, by flushing them away into the rivers via rainfall or irrigation runoff, or leaching them through the soil into groundwater bodies (Beitz et al., 2012). Pesticides applied in urban areas may also reach the aquatic environment after incomplete removal in wastewater treatment plants (WWTPs) (Köck-Schulmeyer et al., 2013; Rousis et al., 2017). Once applied or released in the environment, these compounds can be transformed by different natural processes (photodegradation, hydrolysis, oxidation, biotransformation). These processes hardly mineralize the parent compounds and, consequently, a wide spectrum of transformation products (TPs) are formed, some of them being even more toxic than the corresponding parent compound (Andreu and Picó, 2004; Richardson and Ternes, 2014).

To reduce contamination by pesticides and TPs in the environment and minimize their impact on aquatic organisms and human health, the European Commission has established guidelines that influence the selection and application of pesticides, as well as maximum allowable concentrations (MAC) in both surface water and groundwater. The Groundwater Directive (EC, 2006a) sets maximum limits of 0.1 µg/L for individual pesticides and TPs and 0.5 µg/L for total pesticides in groundwater to preserve its quality, in line with the limits set in the Drinking Water Directive for waters intended for human consumption (EC, 1998). In surface water, the Directive 2013/39/EU (EC, 2013) establishes MACs for up to 45 priority substances, including 24 pesticides or biocides, in inland and other surface waters as well as in biota. Furthermore, five neonicotinoid pesticides, the carbamate methiocarb, and the semicarbamazone metaflumizone are currently included in the Watch List of substances for Union-wide monitoring in the field of water policy (European Decision 2018/840 (EC, 2018)).

To meet European requirements, it is necessary to use analytical techniques that allow the monitoring of pesticide residues in surface and groundwater with high selectivity and sensitivity, such as liquid chromatography (LC) coupled to mass spectrometry (MS or $MS^2$) (Hernández et al., 2005; Picó and Barceló, 2015). However, to measure the low concentrations at which some pesticides are present and toxic in water, a sample enrichment step is still required. Solid phase extraction (SPE) is nowadays considered the method of choice for this purpose (Pérez-Fernández et al., 2017). SPE can be performed in off-line mode, but the fully automated on-line approach has become increasingly attractive as it requires minimum intervention of the operator, which results in reduced sample processing time and improved reproducibility and accuracy of the results. Moreover, it grants high sample throughput and high sensitivity using low sample amounts, which becomes very relevant in case of required sample shipment and/or small storage space (Rossi and Zhang, 2000; Singer et al., 2010). In spite of these advantages, only few methodologies published for the analysis of polar pesticides in water are fully automated (Camilleri et al., 2015; Hurtado-Sánchez et al., 2013; Mann et al., 2016; Quintana et al., 2019; Rubirola et al., 2017; Singer et al., 2010).

In this context, and with the final aim of advancing knowledge on the analysis and monitoring of regulated and non-regulated medium to highly polar pesticides in water bodies, the present work focused on developing and validating a fast and simple analytical methodology based on on-line SPE-LC-MS/MS to determine 51 pesticides in environmental waters at pg/L or ng/L levels in a single run. The selected pesticides belong to different chemical classes and the list includes pesticides of major concern included in the priority substances list or the Watch List, some of their transformation products, and pesticides commonly applied in Spain, and particularly, in Catalonia. To the best of our knowledge, this work provides for the first time validation figures for the analysis of 10 pesticides and TPs (i.e., azinphos ethyl, azinphos-methyl oxon, dichlorvos, diflufenican, fenthion oxon, fenthion oxon sulfone, fenthion oxon sulfoxide, fenthion sulfone, fenthion sulfoxide, and oxadiazon) in water samples with the use of an on-line SPE-LC-MS/MS approach.

As a part of the validation process and to fulfil the overall aim of our work, the developed method was applied to the analysis of surface water and groundwater samples collected in two agriculture-impacted areas of Catalonia (NE Spain), to evaluate the occurrence and fate of the target pesticides and compliance of these water bodies with the current EQS. The results obtained were also used to assess the potential environmental risk that the pesticides found may pose for aquatic organisms in these areas.

## 2. Materials and methods

### *2.1 Standards and solvents*

High purity (>96%) standards of the 51 pesticides selected as target analytes and stable isotope-labeled (SIL) analogs for 45 of them were purchased from Fluka (Honeywell Specialty Chemicals Seelze GmbH, Germany), Toronto Research Chemicals (North York, ON, Canada), Cambridge Isotope Laboratories (Tewksbury, MA, USA), Sigma Aldrich (Merck KGaA,

Darmstadt, Germany) or Dr. Ehrenstorfer (LGC Standards, Teddington, UK). The list of the target analytes and their main physical-chemical properties is provided in Table 1.

Stock standard solutions of the individual analytes and SIL standards were prepared in methanol (MeOH), except in the case of simazine and its SIL analog that was prepared in dimethyl sulfoxide. All stock individual solutions (1000 µg/mL) were stored in amber glass bottles in the dark at -20 °C. Working standard solutions containing all analytes were prepared by appropriate dilution of the stock individual standard solutions at different concentrations (0.5 to 2000 ng/mL) in MeOH. A MeOH-based solution containing the mixture of SIL standards at a concentration of 1000 ng/mL was also prepared. These mixtures were used to prepare the aqueous standard solutions that defined calibration curves and in the validation studies. Pesticides-grade solvents MeOH, acetonitrile (ACN), and LC-grade water were supplied by Merck (Darmstadt, Germany).

**Table 1.** Target pesticides, main physical-chemical properties, and current legislative status.

| Analyte | Chemical class | Formula‡ | MM (g mol$^{-1}$)‡ | Solubility (mg L$^{-1}$) ‡ | $K_{oc}$ (mL g$^{-1}$) ‡ | $K_{ow}$ logP‡ | GUS‡ | DT50‡ (days) | Legislative status‡ | Currently used in Spain‡ | EQS$^{\Omega}$ (µg/L) | Method LODs$^{\varepsilon}$ (ng/L) |
|---|---|---|---|---|---|---|---|---|---|---|---|---|
| **2,4-D** | Alkylchlorophenoxy | $C_8H_6Cl_2O_3$ | 221.04 | 24300 | 39 | -0.82 | 3.82 | 7.7 | ✓ | ✓ | | |
| **Acetamiprid**$^{\varepsilon}$ | Neonicotinoid | $C_{10}H_{11}ClN_4$ | 222.67 | 2950 | 200 | 0.80 | 0.94 | 4.7 | ✓ | ✓ | | 8.3 |
| **Alachlor**$^{\times}$ | Chloroacetamide | $C_{14}H_{20}ClNO_2$ | 269.77 | 240 | 335 | 3.09 | 0.8 | - | X | | 0.7 | |
| **Atrazine**$^{\times}$ | Triazine | $C_8H_{14}ClN_5$ | 215.68 | 35 | 100 | 2.70 | 2.57 | - | X | | 2 | |
| **Azinphos-ethyl** | Organophosphate | $C_{12}H_{16}N_3O_3PS_2$ | 345.38 | 4.5 | 1500 | 3.18 | 1.4 | - | X | | | |
| **Azinphos-methyl** | Organophosphate | $C_{10}H_{12}N_3O_3PS_2$ | 317.32 | 28 | 1112 | 2.96 | 1.42 | - | X | | | |
| **Azinphos-methyl-oxon** | Metabolite | $C_{10}H_{12}N_3O_4PS$ | 301.26 | 2604* | 10 * | 0.77* | - | - | - | | | |
| **Bentazone** | Benzothiazinone | $C_{10}H_{12}N_2O_3S$ | 240.30 | 7112 | 55 | -0.46 | 1.95 | 80 | ✓ | ✓ | | |
| **Bromoxynil** | Hydroxybenzonitrile | $Br_2C_6H_2(OH)CN$ | 276.90 | 38000 | 302 | 0.27 | 1.71 | 13 | ✓ | ✓ | | |
| **Chlorfenvinphos**$^{\times}$ | Organophosphate | $C_{12}H_{14}Cl_3O_4P$ | 359.60 | 145 | 680 | 3.80 | 1.72 | 7 | X | | 0.3 | |
| **Chlorpyrifos**$^{\times}$ | Organophosphate | $C_9H_{11}Cl_3NO_3PS$ | 350.58 | 1.05 | 5509 | 4.70 | 0.58 | 5 | ✓ | ✓ | 0.1 | |
| **Chlortoluron** | Phenylurea | $C_{10}H_{13}ClN_2O$ | 212.68 | 74 | 196 | 2.50 | 2.62 | 42 | ✓ | ✓ | | |
| **Cyanazine** | Triazine | $C_9H_{13}ClN_6$ | 240.69 | 171 | 190 | 2.10 | 2.07 | - | X | | | |
| **Clothianidin**$^{\varepsilon}$ | Neonicotinoid | $C_6H_8ClN_5O_2S$ | 249.68 | 340 | 123 | 0.90 | 3.74 | 40.3 | X | ✓ | | 8.3 |
| **Deisopropylatrazine** | Metabolite | $C_5H_8ClN_5$ | 173.60 | 980 | 130 | 1.15 | - | - | - | | | |
| **Desethylatrazine** | Metabolite | $C_6H_{10}ClN_5$ | 187.63 | 2700 | 110 | 1.51 | 4.37 | - | - | | | |
| **Diazinon** | Organophosphate | $C_{12}H_{21}N_2O_3PS$ | 304.35 | 60 | 609 | 3.69 | 1.51 | 4.3 | X | | | |
| **Dichlorvos**$^{\times}$ | Organophosphate | $C_4H_7Cl_2O_4P$ | 220.98 | 18000 | 50 | 1.90 | 0.69 | - | X | | $7 \times 10^{-4}$ | |
| **Diflufenican** | Carboxamide | $C_{19}H_{11}F_5N_2O_2$ | 394.29 | 0.05 | 5504 | 4.20 | 1.19 | - | ✓ | ✓ | | |
| **Dimethoate** | Organophosphate | $C_5H_{12}NO_3PS_2$ | 229.26 | 25900 | 25* | 0.75 | 2.18 | 12.6 | X | ✓$^{\infty}$ | | |
| **Diuron**$^{\times}$ | Phenylurea | $C_9H_{10}Cl_2N_2O$ | 233.09 | 35.6 | 680 | 2.87 | 2.65 | 8.8 | ✓ | ✓ | 1.8 | |
| **Fenitrothion** | Organophosphate | $C_9H_{12}NO_5PS$ | 277.23 | 19 | 2000 | 3.32 | 0.48 | 1.1 | X | ✓ | | |
| **Fenitrothion oxon** | Metabolite | $C_9H_{12}NO_6P$ | 261.17* | 301 * | 21* | 1.69* | - | - | - | | | |
| **Fenthion** | Organophosphate | $C_{10}H_{15}O_3PS_2$ | 278.33 | 4.2 | 1500 | 4.84 | 1.26 | - | X | | | |
| **Fenthion oxon** | Metabolite | $C_{10}H_{15}O_4PS$ | 262.26* | 213.5* | 57 * | 2.31* | - | - | - | | | |
| **Fenthion oxon sulfone** | Metabolite | $C_{10}H_{15}O_6PS$ | 294.03* | 7602* | 13* | 0.28* | - | - | - | | | |
| **Fenthion oxon sulfoxide** | Metabolite | $C_{10}H_{15}O_5PS$ | 278.26* | 1222* | 11* | 0.15* | - | - | - | | | |
| **Fenthion sulfone** | Metabolite | $C_{10}H_{15}O_5PS_2$ | 310.33* | 190.4* | 235 | 2.05* | - | - | - | | | |
| **Fenthion sulfoxide** | Metabolite | $C_{10}H_{15}O_4PS_2$ | 294.33* | 3.72* | 183 | 1.92* | - | - | - | | | |
| **Fluroxypyr** | Pyridine compound | $C_7H_5Cl_2FN_2O_3$ | 255.03 | 6500 | 10* | 0.04 | 3.7 | 10.5 | ✓ | ✓ | | |
| **Imidacloprid**[ε] | Neonicotinoid | $C_9H_{10}ClN_5O_2$ | 255.66 | 610 | 6719 | 0.57 | 3.69 | 30 | ✓ | ✓ | | 8.3 |
| **Irgarol**[×] | Triazine | $C_{11}H_{19}N_5S$ | 253.37 | 7 | 1569 | 3.95 | - | - | X | | 0.016 | |
| **Isoproturon**[×] | Phenyluera | $C_{12}H_{18}N_2O$ | 206.28 | 70.2 | 251* | 2.5 | 2.61 | 40 | X | ✓ | 1 | |
| **Linuron** | Phenyluera | $C_9H_{10}Cl_2N_2O_2$ | 249.09 | 63.8 | 843 | 3 | 2.11 | 13 | X | ✓ | | |
| **Malaoxon** | Metabolite | $C_{10}H_{19}O_7PS$ | 314.29* | 7500* | 4650* | 0.52* | - | - | - | | | |
| **Malathion** | Organophosphate | $C_{10}H_{19}O_6PS_2$ | 330.36 | 148 | 1800 | 2.75 | 0.00 | 0.4 | ✓ | ✓ | | |
| **MCPA** | Organophosphate | $C_9H_9ClO_3$ | 200.62 | 29390 | 29* | -0.81 | 2.98 | 13.5 | ✓ | ✓ | | |
| **Mecoprop** | Aryloxyalkanoic acid | $C_{10}H_{11}ClO_3$ | 214.65 | 250000 | 47 | -0.19 | 2.29 | 37 | X | ✓ | | |
| **Methiocarb**[ε] | Carbamate | $C_{11}H_{15}NO_2S$ | 225.31 | 27 | 182* | 3.18 | 1.82 | 1.6 | ✓ | ✓ | | 2 |
| **Metolachlor** | Chloroacetamide | $C_{15}H_{22}ClNO_2$ | 283.80 | 530 | 120 | 3.40 | 2.36 | 88 | X | | | |
| **Molinate** | Thiocarbamate | $C_9H_{17}NOS$ | 187.30 | 1100 | 190 | 2.86 | 1.89 | 4 | X | ✓ | | |
| **Pendimethalin** | Dinitroaniline | $C_{13}H_{19}N_3O_4$ | 281.31 | 0.33 | 17491 | 5.40 | -0.28 | 4 | ✓ | ✓ | | |
| **Propanil** | Anilide | $C_9H_9Cl_2NO$ | 218.08 | 95 | 149 | 2.29 | -0.51 | 1.2 | X | ✓ | | |
| **Quinoxyfen**[×] | Quinoline | $C_{15}H_8Cl_2FNO$ | 308.13 | 0.05 | 23° | 4.66 | -0.8 | 5 | ✓ | ✓ | 2.7 | |
| **Simazine**[×] | Triazine | $C_7H_{12}ClN_5$ | 201.66 | 5 | 130 | 2.30 | 2.2 | 46 | X | ✓ | 4 | |
| **Terbuthylazine** | Triazine | $C_9H_{16}ClN_5$ | 229.71 | 6.6 | 329* | 3.40 | 2.19 | 6 | ✓ | ✓ | | |
| **Terbutryn**[×] | Triazine | $C_{10}H_{19}N_5S$ | 241.36 | 25 | 2432 | 3.66 | 2.21 | 27 | X | | 0.34 | |
| **Thiacloprid**[ε] | Neonicotinoid | $C_{10}H_9ClN_4S$ | 252.72 | 184 | 615° | 1.26 | 1.1 | 1000 | ✓ | | | 8.3 |
| **Thiamethoxam**[ε] | Neonicotinoid | $C_8H_{10}ClN_5O_3S$ | 291.71 | 4100 | 56 | -0.13 | 3.58 | 30.6 | X | ✓ | | 8.3 |
| **Thifensulfuron methyl** | Sulfonylurea | $C_{12}H_{13}N_5O_6S_2$ | 387.39 | 54.1 | 28 | -1.65 | 3.05 | 22 | ✓ | ✓ | | |
| **Triallate** | Thiocarbamate | $C_{10}H_6Cl_3NOS$ | 304.7 | 4.1 | 3034 | 4.06 | 0.61 | 104 | ✓ | ✓ | | |

[×] Compound included in the list of priority substances. EC Directive 2013/39/EU of the European Parliament and of the Council of 12 August 2013 amending Directives 2000/60/EC and 2008/105/EC as regard priority substances in the field of water policy. Retrieved from: https://goo.gl/diHn8W.

[ε] Compound included in the European Watch List and corresponding maximum acceptable method detection limit (ng/L). EC Commission Implementing Decision (EU) 2018/840 of 5 June 2018 establishing a watch list of substances for Union-wide monitoring in the field of water policy pursuant to Directive 2008/105/EC of the European Parliament and of the Council and repealing Commission Implementing Decision (EU) 2015/495 (notified under document C(2018) 3362). Retrieved from: https://goo.gl/nR4ezg.

‡ The PPDB, Pesticide Properties Database. http://sitem.herts.ac.uk/aeru/footprint/index2.htm. - Lewis, K.A., Tzilivakis, J., Warner, D. and Green, A. (2016). An international database for pesticide risk assessments and management. Human and Ecological Risk Assessment: An International Journal, 22(4), 1050-1064.

* Data estimated using the US Environmental Protection Agency EPISuite™ http://www.Chemspider.com.

° Kegley, S.E., Hill, B.R., Orme S., Choi A.H., PAN Pesticide Database, Pesticide Action Network, North America (Oakland, CA, 2016), http://www.pesticideinfo.org.

^ Calculated using the mathematical formula: GUS = log10 (half-life) x [4 - log10 (Koc)].

Ω Environmental Quality Standards (EQS) for priority substances in surface waters.

∞ Commission implementing regulation (EU) 2019/1090 of 26 June 2019 concerning the non-renewal of approval of the active substance dimethoate. Member States shall withdraw authorizations for plant protection products containing dimethoate as active substance by 17 January 2020 at the latest.

MM: molecular mass; Solubility: solubility in water at 20 $^{o}$C; $K_{oc}$: organic carbon partition coefficient; $K_{ow}$: octanol-water partition coefficient; GUS: leaching potential index; DT50: biodegradability, water phase only, expressed as *half-life* in days; Legislative status: ✓ approved, **X** not approved.

## 2.2 On-line solid-phase extraction

On-line SPE of the water samples was performed with a commercial Prospekt-2 system (Spark Holland, Emmen, The Netherlands) connected in series with the LC-MS/MS instrument. Before automated on-line SPE and analysis, the water sample was fortified at a concentration of 200 ng/L with the mixture of SIL compounds that will be used as surrogate standards and centrifuged at g-force of 2500 xg (3500 rpm) and room temperature for 10 min to remove suspended particles (centrifuge 5810 R, Eppendorf Ibérica, Spain). Then, 5 mL of the sample, calibration solution and/or blank was delivered using a 2 mL high-pressure syringe onto a previously conditioned CHROspe cartridge Polymer DVB (divinylbenzene polymer, 10 mm x 2 mm i.d., 25-35 µm particle size) (Axel Semrau GmbH & Co. KG, Srockhövel, Germany) at a flow rate of 1 mL/min. Conditioning of the cartridge was also performed via the high pressure dispenser (HPD) unit with 1 mL of ACN and 1 mL of LC-grade water (5 mL/min). Upon sample loading, the cartridge was washed with 1 mL of LC-grade water to complete sample transfer and remove highly polar components of the matrix and the analytes were eluted with the LC mobile phase onto the LC analytical column. The system configuration allows the preconcentration of the next sample in a batch while the LC-MS analysis of the previously extracted sample takes place. The entire system was controlled through SparkLink Version 3.10 (Spark Holland).

## 2.3 LC-MS/MS analysis

LC-MS/MS analyses were performed using a 1525 binary HPLC pump connected in series with the Prospekt-2 system and a TQD triple-quadrupole mass spectrometer equipped with an electrospray (ESI) interface (Waters, Milford, MA, USA).

LC separation was carried out with a Purospher® STAR RP-18 end-capped column (100 mm x 2 mm i.d., 5 µm particle size) preceded by a guard column (4 mm x 4 mm i.d., 5 µm) of the same packing material (Merck, Darmstadt, Germany), and a linear gradient of ACN and water as mobile phase at a flow rate of 0.2 mL/min. The gradient started with an ACN composition of 10% that was increased to 50% in 5 min, to 80% in the next 20 min, and 100% in the following 6 min. Then, the chromatographic column was reequilibrated with the mobile phase initial conditions in the following 9 min. In total, the analysis time, including the sample extraction step, was 40 min.

MS/MS detection was performed in the selected reaction monitoring (SRM) mode, recording one SRM transition per SIL compound and two SRM transitions per target analyte, with the first one and more abundant being used for quantification and the second one for confirmation. A total of 146 SRM transitions was acquired in six separate retention windows, to maximize the sensitivity of the MS instrument (Figure 1). The ESI interface was operated in both positive (PI) and negative (NI) ionization modes according to the preferential ionization mode of the target analytes (43 were analyzed in PI and 8 in NI). Table 2 summarizes the optimum SRM transitions and ionization conditions for each selected analyte. Other specific optimized MS conditions were as follows: capillarity voltage, 3.5 kV; extractor voltage, 3 V; RF lens voltage, 1.8 V; source temperature, 150 °C; desolvation temperature, 450 °C. Nitrogen was used as cone gas (flow, 30 L/Hr) and desolvation gas (flow, 680 L/Hr); and argon was used as collision gas (flow, 0.19 mL/min). MassLynx 4.1 software from Waters was used to perform instrument control, data acquisition, and quantification.

**Table 2.** On-line SPE-LC-MS/MS conditions for the analysis of the 51 investigated pesticides and SIL analogs.

| Analyte | Retention time (min) | SRMs (m/z) Precursor ion>product ion | Cone (V) | Collision Energy (eV) | SRM ratio (SRM1/SRM2) |
|---|---|---|---|---|---|
| ***Negative ESI mode*** | | | | | |
| Bromoxynil | 7.5 | 276>81/276>79 | 40 | 20/30 | 16.6 |
| *Bromoxynil $^{13}C_6$* | | *282/81* | *40* | *20* | |
| Bentazone | 7.5 | 239>132/239>197 | 30 | 25/20 | 2.8 |
| *Bentazone $d_6$* | | *245>132* | *35* | *25* | |
| Fluroxypyr[α] | 7.8 | 253>195/255>197 | 15 | 10/10 | 1.8 |
| 2,4-D | 7.9 | 219>161/219>125 | 20 | 15/25 | 15.4 |
| *2,4-D $d_3$* | | *224>127* | *20* | *25* | |
| MCPA | 7.9 | 199>141/201>143 | 25 | 10/10 | 2.7 |
| *MCPA $d_3$* | | *204>146* | *25* | *20* | |
| Mecoprop | 7.9 | 213>141/213>71 | 25 | 10/10 | 10.2 |
| *Mecoprop $d_3$* | | *218>146* | *25* | *15* | |
| Propanil | 14.9 | 216>160/218>162 | 25 | 20/20 | 1.4 |
| *Propanil $d_5$* | | *221>161* | *30* | *15* | |
| Fenitrothion | 19.8 | 262>152/262>122 | 25 | 20/30 | 9.6 |
| *Fenitrothion $d_6$* | | *265>152* | *30* | *15* | |
| ***Positive ESI mode*** | | | | | |
| Thifensulfuron methyl | 7.3 | 388>167/388>141 | 25 | 15/20 | 6.4 |
| ***Thifensulfuron methyl*** *$d_3$* | | *391>167* | *20* | *15* | |
| Desethylatrazine (DEA) | 7.4 | 188>146/188>79 | 30 | 15/25 | 11.5 |
| ***Desethylatrazine*** *$d_6$* | | 194>147 | 25 | 20 | |
| Fenthion oxon sulfoxide[β] | 8.2 | 279>264/279>104 | 35 | 20/25 | 1.3 |
| Deisopropilatrazine (DIA) | 8.5 | 174>132/174>104 | 30 | 20/25 | 1.1 |
| ***Deisopropilatrazine*** *$d_5$* | | 179>101 | 40 | 20 | |
| **Thiamethoxam** | 8.9 | 292>211/292>181 | 25 | 15/20 | 3.9 |
| ***Thiamethoxam*** *$d_3$* | | 295>214 | 30 | 15 | |
| Clothianidin | 8.9 | 250>169/250>132 | 20 | 15/15 | 2.1 |
| ***Clothianidin*** *$d_3$* | | 254>172 | 20 | 10 | |
| Imidacloprid | 9.0 | 256>175/256>209 | 25 | 20/15 | 1.3 |
| ***Imidacloprid*** *$d_5$* | | 261>214 | 25 | 25 | |
| Acetamiprid | 9.2 | 223>126/223>56 | 15 | 15/20 | 1.9 |
| ***Acetamiprid*** *$d_3$* | | 227>126 | 35 | 20 | |
| Dimethoate | 9.6 | 230>199/230>125 | 25 | 15/15 | 3.5 |
| ***Dimethoate*** *$d_6$* | | 236>131 | 30 | 15 | |
| Azinphos methyl oxon[γ] | 9.7 | 324>132/324>148 | 40 | 20/15 | 18.8 |
| Thiacloprid | 10.1 | 253>126/253>90 | 25 | 25/40 | 4.3 |
| ***Thiacloprid*** *$d_4$* | | 257>126 | 35 | 15 | |
| Dichlorvos | 10.8 | 221>109/223>109 | 30 | 25/25 | 1.1 |
| ***Dichlorvos*** *$d_6$* | | 227>115 | 30 | 25 | |
| Simazine | 10.8 | 202>124/202>71 | 30 | 25/20 | 3.0 |
| ***Simazine*** *$d_{10}$* | | 212>137 | 35 | 15 | |
| Cyanazine | 10.9 | 241>214/241>174 | 30 | 15/20 | 13.2 |
| ***Cyanazine*** *$d_5$* | | 246>219 | 30 | 20 | |
| Malaoxon (MOX)[δ] | 11.0 | 315>99/315>127 | 20 | 25/15 | 1.4 |
| Fenthion oxon sulfone | 11.1 | 295>109/295>217 | 40 | 40/30 | 27.8 |
| ***Fenthion oxon sulfone*** *$d_3$* | | 298>104 | 35 | 25 | |
| Fenthion sulfoxide | 11.2 | 295>109/295>125 | 40 | 30/35 | 4.3 |
| ***Fenthion sulfoxide*** *$d_6$* | | 301>108 | 35 | 30 | |
| Fenitrothion oxon | 11.5 | 262>104/262>216 | 30 | 20/20 | 4.6 |
| ***Fenitrothion oxon*** *$d_6$* | | 268>106 | 30 | 25 | |
| Chlortoluron | 11.9 | 213>72/213>140 | 25 | 15/30 | 47.3 |
| ***Chlortoluron*** *$d_6$* | | 219>78 | 35 | 15 | |
| Isoproturon | 12.3 | 207>165/207>72 | 35 | 15/20 | 9.3 |
| ***Isoproturon*** *$d_6$* | | 213>171 | 30 | 20 | |
| Atrazine | 12.4 | 216>174/216>132 | 35 | 15/20 | 6.5 |
| ***Atrazine*** *$d_5$* | | 221>179 | 35 | 15 | |
| Diuron | 12.8 | 233>72/235>72 | 25 | 15/15 | 1.5 |
| ***Diuron*** $d_6$ | | 239>78 | 25 | 25 | |
| Fenthion oxon | 13.0 | 263>231/263>216 | 35 | 20/25 | 1.2 |
| ***Fenthion oxon*** $d_3$ | | 266>234 | 30 | 15 | |
| Fenthion sulfone | 14.2 | 311>125/311>109 | 35 | 30/40 | 2.7 |
| ***Fenthion sulfone*** $d_6$ | | 317>115 | 45 | 30 | |
| Terbuthylazine | 15.2 | 230>174/230>96 | 25 | 15/25 | 4.3 |
| ***Terbuthylazine*** $d_5$ | | 235>179 | 30 | 20 | |
| Methiocarb | 15.4 | 226>169/226>121 | 20 | 10/20 | 2.2 |
| ***Methiocarb*** $d_3$ | | 229>169 | 25 | 10 | |
| Linuron | 16.2 | 249>160/249>182 | 25 | 15/15 | 1.3 |
| ***Linuron*** $d_6$ | | 255>185 | 30 | 15 | |
| Azinphos methyl | 16.5 | 318>132/318>105 | 15 | 20/30 | 2.1 |
| ***Azinphos methyl*** $d_6$ | | 324>132 | 35 | 15 | |
| Molinate[ε] | 17.4 | 188>126/188>83 | 30 | 15/20 | 1.5 |
| Terbutryn | 17.6 | 242>71/242>91 | 30 | 30/30 | 1.4 |
| ***Terbutryn*** $d_5$ | | 247>191 | 35 | 20 | |
| Irgarol | 17.7 | 254>108/254>125 | 35 | 30/25 | 2.9 |
| ***Irgarol*** $d_9$ | | 283>199 | 40 | 20 | |
| Metolachlor | 18.3 | 284>176/284>73 | 25 | 25/25 | 6.5 |
| ***Metolachlor*** $d_{11}$ | | 295>263 | 25 | 15 | |
| Alachlor | 18.5 | 270>238/270>162 | 30 | 15/15 | 1.2 |
| ***Alachlor d*** $_{13}$ | | 283>251 | 15 | 10 | |
| Malathion | 18.9 | 353>195/353>227 | 30 | 15/15 | 3.5 |
| ***Malathion*** $d_{10}$ | | 363>205 | 35 | 15 | |
| Chlorfenvinphos (CFP) | 19.2 | 359>155/359>170 | 25 | 15/40 | 1.4 |
| ***Chlorfenvinphos*** $d_{10}$ | | 369>101 | 25 | 30 | |
| Azinphos ethyl | 19.8 | 346>132/346>104 | 15 | 20/35 | 2.9 |
| ***Azinphos ethyl*** $d_{10}$ | | 356>132 | 15 | 20 | |
| Diazinon | 21.9 | 305>153/305>97 | 35 | 20/30 | 1.7 |
| ***Diazinon*** $d_{10}$ | | 315>170 | 35 | 20 | |
| Diflufenican | 25.3 | 395>266/395>246 | 45 | 30/25 | 7.0 |
| ***Diflufenican*** $d_5$ | | 398>268 | 35 | 25 | |
| Oxadiazon | 28.3 | 345>220/345>177 | 35 | 30/20 | 1.1 |
| ***Oxadiazon*** $d_7$ | | 352>221 | 35 | 20 | |
| Quinoxyfen[ζ] | 28.6 | 309>245/309>150 | 50 | 30/30 | 1.1 |
| Pendimethalin | 29.2 | 282>212/282>194 | 15 | 10/15 | 27.4 |
| ***Pendimethalin*** $d_5$ | | 287>213 | 20 | 10 | |
| Chlorpyrifos (CPF) | 29.5 | 352>97/352>200 | 20 | 30/20 | 59.1 |
| ***Chlorpyrifos*** $d_{10}$ | | 362>131 | 25 | 20 | |
| Triallate | 29.6 | 304>86/304>143 | 25 | 15/30 | 3.8 |
| ***Triallate*** $^{13}C_6$ | | 310>89 | 30 | 15 | |

[α]Compound quantified using mecoprop $d_3$ as surrogate standard.
[β]Compound quantified using thiamethoxam $d_3$ as surrogate standard.
[γ]Compound quantified using fenthion sulfoxide $d_6$ as surrogate standard.
[δ]Compound quantified using chlortoluron $d_6$ as surrogate standard.
[ε]Compound quantified using linuron $d_6$ as surrogate standard.
[ζ]Compound quantified using chlorpyrifos $d_{10}$ as surrogate standard.

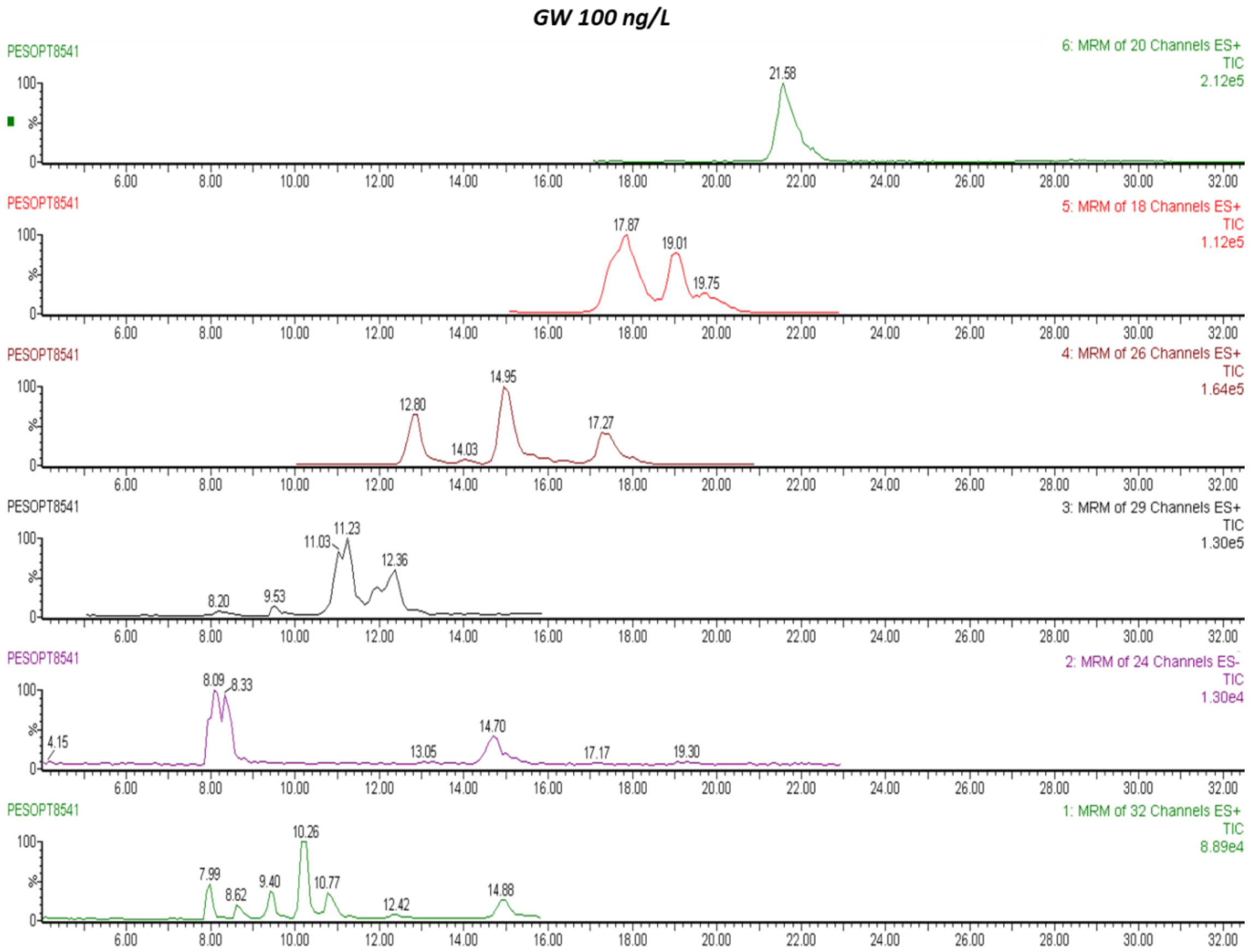

**Figure 1.** Total Ion Current (TIC) chromatograms obtained from the analysis of a fortified groundwater sample (100 ng/L) showing the acquisition of the 146 SRM transitions set for determination of the 51 target pesticides and their 45 SIL analogs in six different acquisition windows along the analytical run.

## *2.4 Method performance*

The analytical method was validated in terms of linearity, accuracy, precision, sensitivity, and matrix effects, in both surface water and groundwater. The validation in groundwater was carried out using a pooled sample of groundwater from various aquifers located in Catalonia (NE Spain). In the case of surface water, a pooled sample of water from three Catalonian rivers, namely, Segre, Llobregat, and Tordera, was used.

Eleven calibration solutions within the concentration range 0.5-2000 ng/L, constructed after appropriate dilution of the working standard solutions in LC-grade water were used to evaluate method linearity. Quantification was done using an isotope dilution approach, i.e., considering the ratio between the peak area of each analyte and that of its corresponding SIL analog, except in the case of six compounds, for which SIL analogs were not available. These compounds were quantified using SIL standards presenting similar structure, retention time, and/or recoveries. Method linearity was expressed with the coefficient of determination ($r^2$) of the weighted linear regression model obtained for each analyte. $1/x^2$ was used as a weighting factor to reduce the influence of the high concentration data points in the model.

Accuracy and precision of the method in LC-grade water, surface, and groundwater were appraised with the analyte recovery and its repeatability after n=5 replicated analyses of each matrix fortified at three different concentration levels (10 ng/L, 100 ng/L, and 1000 ng/L). Background concentration levels of each target pesticide in each matrix were taken into account in the calculations.

The method sensitivity was evaluated through the calculation of limits of detection (LOD), limits of quantification (LOQ), and limits of determination (LODet). LOD and LOQ were experimentally estimated from the analysis of the water matrices fortified at the lowest level (10 ng/L) as the analyte concentration giving a signal to noise ratio of 3 in the case of LOD and 10 in the case of LOQ. LODet coincides with the minimum concentration of a compound that can be quantified (LOQ of SRM1) and confirmed (LOD of SRM2).

To evaluate the matrix effects produced by co-extracted matrix components, analyte peak areas obtained after on-line SPE-LC-MS/MS analysis of surface water and groundwater fortified at 100 ng/L were compared with those obtained after on-line SPE-LC-MS/MS analysis of LC-grade water fortified at equal concentration. Negative matrix effect values occur when the analyte signal in LC-grade water is higher than in fortified surface or groundwater, and indicate ionization suppression effects. On the contrary, positive matrix effect values occur when the analyte signal is higher in surface and groundwater than in LC-grade water, and indicate signal enhancement effects.

### *2.5 Sampling locations and water collection*

The presence of the target pesticides was investigated in the last stretches of the two largest river basins of Catalonia (NE Spain), i.e., the Llobregat and the Ter River basins. Surface water of these two basins is used to supply drinking water to about 4.5 million people in Barcelona and its metropolitan area (Postigo et al., 2018). The Llobregat River is located in an area with an important concentration of industries (e.g., tannery, food products, textile, pulp, and paper industries), and high population density, and thus, with an important demand of water. This river is highly impacted by domestic and industrial wastewater discharges (> 30 wastewater treatment plants) and surface runoff from agricultural areas (González et al., 2012). On the contrary, the Ter River basin is characterized by a low population density and intense agricultural activities (e.g., crops of rice, corn, alfalfa, and apple trees, among others). It also receives the impact of some metallurgic, pulp mill, textile, and tannery industries (Céspedes et al., 2006).

The sampling campaigns were conducted in February 2017 in the Llobregat River and in June 2018 in the Ter River. A total of 11 surface water samples were collected from the Llobregat River, and the same number (6 surface water and 5 groundwater samples) from the Ter River (Figure 2). Grab sampling was done in all surface water locations. Groundwater samples were collected after pumping each well for few minutes (10-20 min) to remove stagnant water, at the minimum flow rate possible, and steady conditions of physical-chemical parameters (i.e. temperature, pH, and conductivity). All samples were collected in amber polyethylene terephthalate (PET) bottles and transported under cool conditions to the laboratory, where they were stored upon arrival at -20 °C in the dark until analysis.

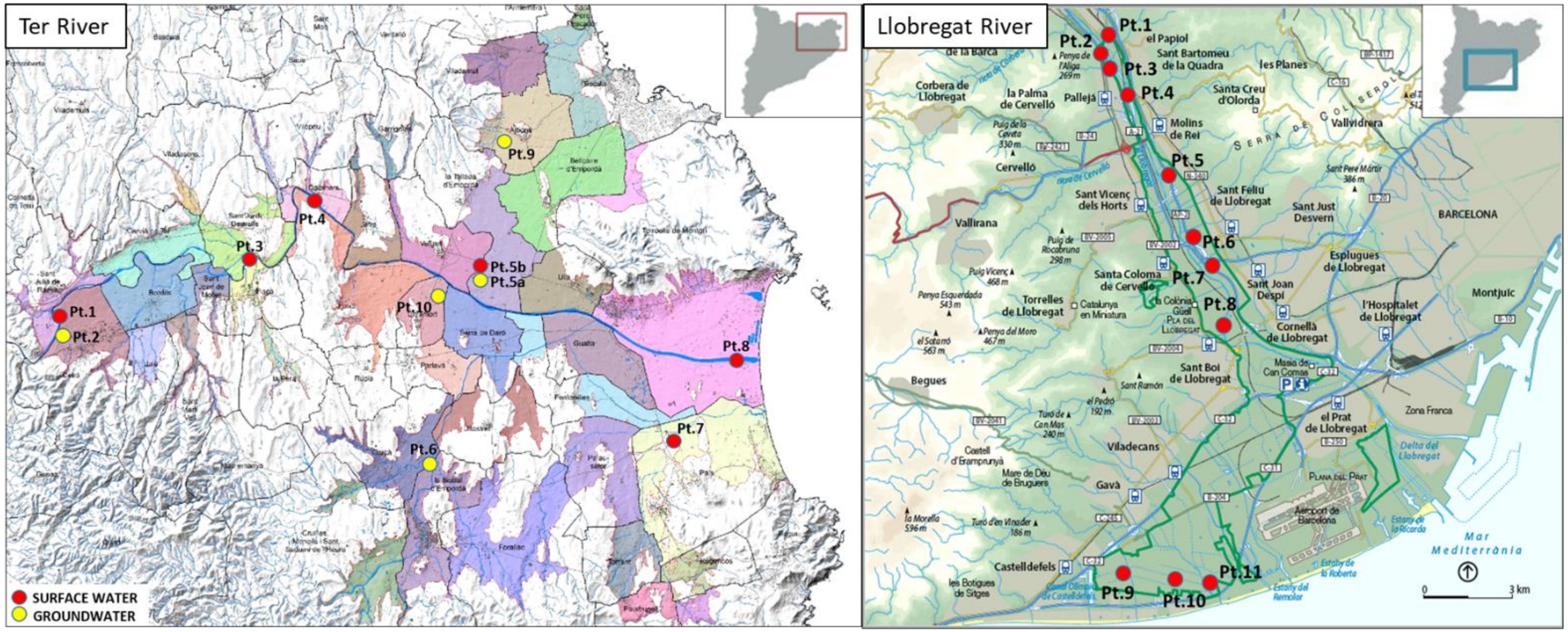

**Figure 2**. Sampling locations in each of the river basins investigated. Figure courtesy of J. Montaner from IRTA (Ter) and V. Sola from CUADLL (Llobregat).

### *2.6 Risk assessment*

The potential environmental risk associated with the pesticides found in the investigated samples was assessed using the hazard quotient (HQ) approach (EPA, 1997), following the equation: HQ=MEC/PNEC. This approach compares the measured environmental concentration (MEC) for each compound with its predicted no-effect concentration (PNEC), i.e., the concentration at which no toxic effects are expected to occur. To assess the worst-case scenario, the maximum pesticide concentration measured in the various investigated samples ($MEC_{max}$) was used as MEC, and the PNEC was the lowest PNEC value provided in the NORMAN Ecotoxicology Database (https://www.norman-network.com/nds/ecotox/) (Dulio and von der OH, 2013). Furthermore, the overall effect of the pesticide mixtures present in the samples was evaluated using an additive model, i.e., adding the individual HQs of the pesticides measured in each sample. It was considered that with an HQ<0.1, no adverse effects are expected for the aquatic organisms; 0.1<HQ<1 means that the risk is low but potential adverse effects cannot be fully dismissed; HQ>1 means that some adverse effects or moderate risk is probable, and with HQ>10 a high risk is anticipated.

Additional risk assessment was conducted by comparing measured concentrations with quality standards in surface and groundwater.

## 3. Results and discussion

### *3.1 Method optimization*

The analytical method developed is based on a methodology previously described for the analysis of 22 pesticides in environmental waters (Köck-Schulmeyer et al., 2014; Köck-Schulmeyer et al., 2013; Postigo et al., 2010). One of the analytical improvements incorporated in this methodology is the expansion of the list of the targeted pesticides with 29 additional medium to highly polar pesticides, including 8 TPs, i.e., acetamiprid, azinphos ethyl, azinphos-methyl, azinphos-methyl oxon, bromoxynil, chlorfenvinphos, chlorpyrifos, clothianidin, dichlorvos, diflufenican, fenitrothion oxon, fenthion oxon, fenthion oxon sulfone, fenthion oxon sulfoxide, fenthion sulfone, fenthion sulfoxide, fluroxypyr, imidacloprid, irgarol, malaoxon, methiocarb, oxadiazon, pendimethalin, quinoxyfen, terbutryn, thiacloprid, thiamethoxam, thifensulfuron methyl, and triallate, and SIL analogs for 24 of them. These pesticides and TPs were selected considering their feasibility for LC-MS analysis, their current use in Spain, and their inclusion in the EU legislation (as a priority, Watch List or banned substances) (EC, 2013; EC, 2018) (Table 1). The current method, unlike the previous one, removes suspended particles by centrifugation instead of filtration. Moreover, this step is conducted after surrogate standard addition to account for pesticides present in the whole matrix and reduce the loss of analytes due to adsorption onto the filters. Compared to the previous method, the analysis time is reduced to half due to the determination of all targeted pesticides and TPs in a single analytical run. This was possible thanks to the use of a generic sorbent for the simultaneous preconcentration of all analytes, and the switch of polarity ionization during MS acquisition.

The optimization of the MS/MS conditions for the detection of the new analytes included in the methodology was performed by on-column injection of individual standard solutions of each compound. Full scan acquisition was used to select the molecular ion and the best ionization mode and optimum declustering potential for its detection, and product ion scan acquisition allowed obtaining the optimum collision energies to register the two most abundant and selective fragment ions (SRM transitions) in each case. MS/MS conditions for each pesticide and SIL analog are provided in Table 2. Up to six time-acquisition windows were established to maximize the acquisition time for each SRM transition and hence improve method sensitivity. An example of the extracted ion chromatograms of the target compounds obtained after on-line SPE-LC-MS/MS analysis of a surface water sample fortified with the targeted pesticides at a concentration of 100 ng/L (500 ng/L for those compounds with LOD above 100 ng/L) is provided in Figure 3.

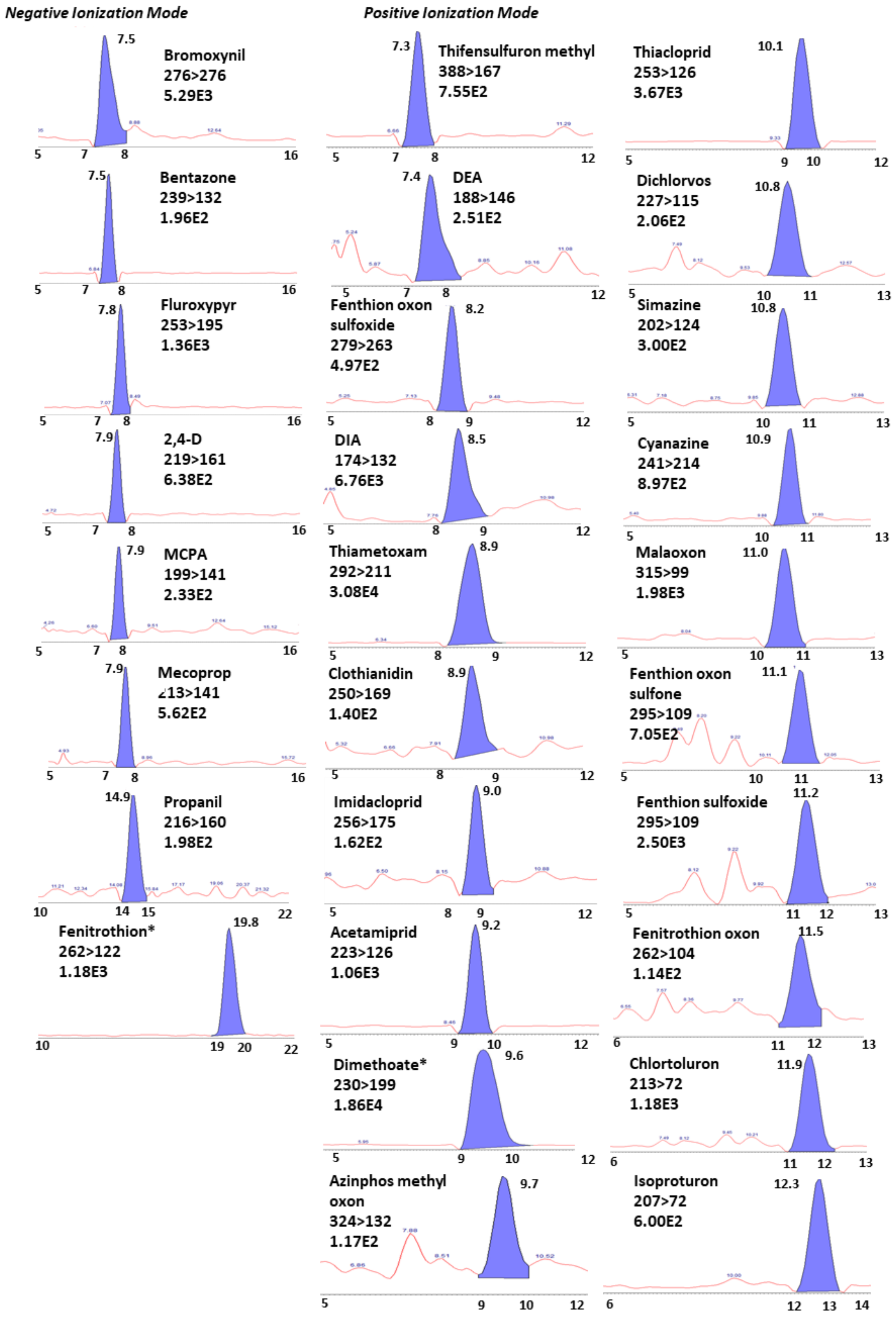

**Figure 3.** Extracted ion chromatograms (XIC) of the target pesticides after on-line SPE-LC-MS/MS analysis of a surface water sample fortified at a concentration of 100 ng/L (or 500 ng/L in the case of those compounds marked with *).

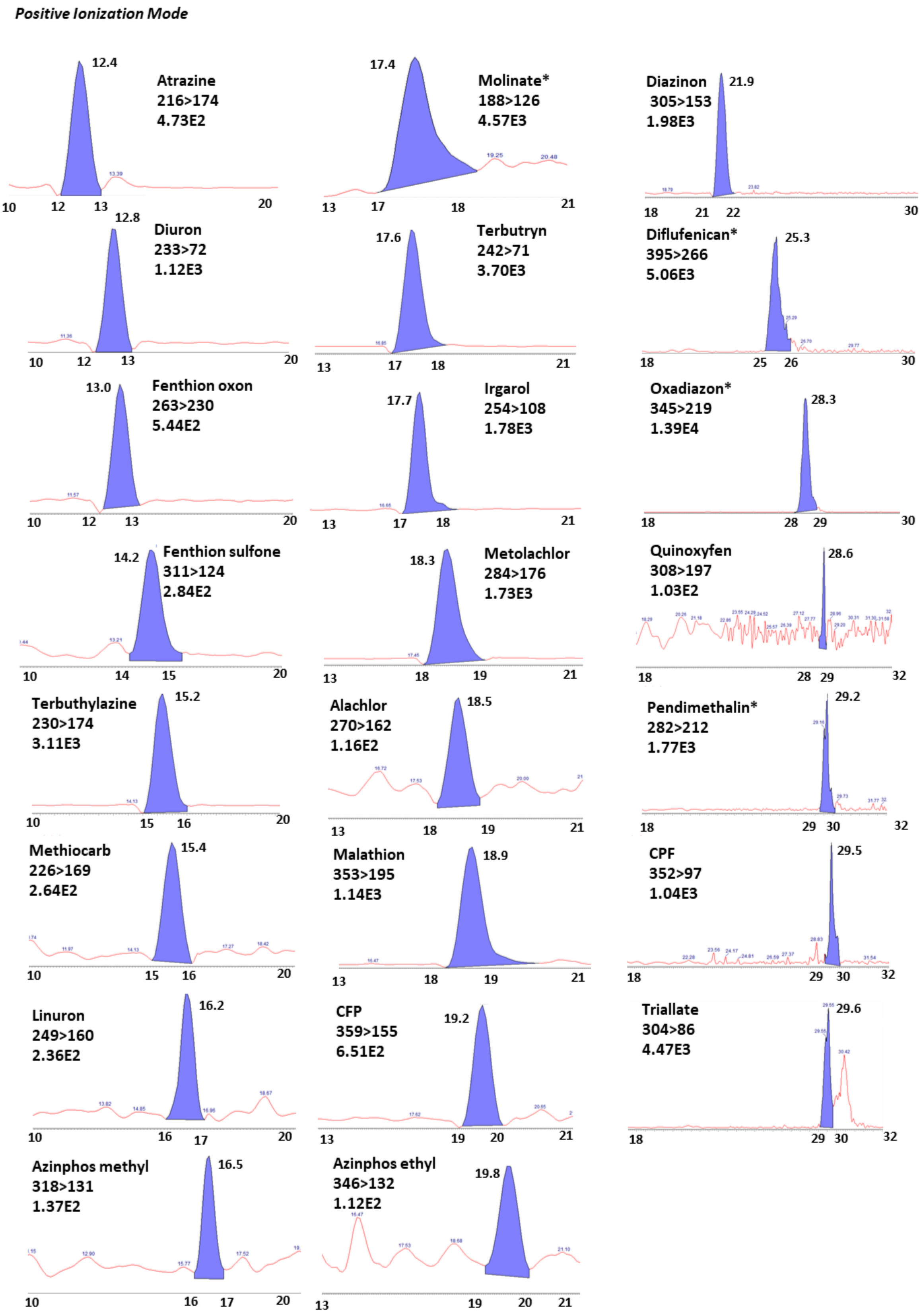

**Figure 3.** (continued).

### *3.2 Method validation*

Tables 3-5 and Figure 4 summarize the method performance in groundwater, surface water, and LC-grade water, in terms of linearity, recovery, repeatability, sensitivity, and matrix effects at the three concentration levels investigated (10 ng/L, 100 ng/L, and 1000 ng/L).

The linearity of the method expanded between 0.5 ng/L and 2000 ng/L for most of the compounds (1000 ng/L was the upper linearity range in the case of clothianidin, dichlorvos, diuron, fenitrothion oxon, fenthion sulfone, fenthion oxon sulfone, malathion, mecoprop, molinate, simazine, and thifensulfuron methyl). The weighted linear regression models presented a coefficient of determination ($r^2$) higher than 0.99 for all compounds except for fenitrothion (0.981) and fenthion sulfone (0.983) (Table 3).

**Table 3.** Method performance in terms of linearity, recovery, repeatability (RSD, relative standard deviation), and sensitivity (limits of detection (LOD) and limits of determination (LODet)) for the target pesticides in surface water and groundwater.

| Analyte | Linearity ($r^2$) | *Groundwater* | | | *Surface water* | | |
|---|---|---|---|---|---|---|---|
| | | Accuracy and precision (100 ng/L) | Sensitivity | | Accuracy and precision (100 ng/L) | Sensitivity | |
| | | Analyte recovery ± RSD(%) | LOD ng/L | LODet ng/L | Analyte recovery ± RSD(%) | LOD ng/L | LODet ng/L |
| **2,4-D** | 0.9947 | 82 ± 10 | 4.0 | 13 | 104 ± 16 | 2.4 | 14 |
| **Acetamiprid** | 0.9900 | 115 ± 13 | 1.4 | 5.4 | 84 ± 7 | 1.0 | 4.0 |
| **Alachlor** | 0.9988 | 81 ± 12 | 2.1 | 7.1 | 116 ± 12 | 6.4 | 16 |
| **Atrazine** | 0.9964 | 93 ± 16 | 1.4 | 4.8 | 93 ± 3 | 1.9 | 6.7 |
| **Azinphos ethyl** | 0.9939 | 96 ± 10 | 2.6 | 8.7 | 118 ± 5 | 3.8 | 9.3 |
| **Azinphos methyl** | 0.9900 | 119 ± 15 | 1.7 | 5.6 | 81 ± 7 | 4.3 | 12 |
| **Azinphos methyl oxon** | 0.9978 | 81 ± 6 | 5.3 | 18 | 88 ± 15 | 15 | 27 |
| **Bentazone** | 0.9918 | 83 ± 9 | 2.5 | 9.3 | 101 ± 5 | 1.9 | 8.8 |
| **Bromoxynil** | 0.9901 | 84 ± 6 | 0.41 | 1.4 | 82 ± 11 | 8.5 | 22 |
| **Chlorfenvinphos** | 0.9988 | 86 ± 13 | 0.50 | 1.7 | 112 ± 4 | 0.59 | 2.9 |
| **Chlorpyrifos** | 0.9913 | 113 ± 13 | 1.7 | 5.6 | 97 ± 6 | 0.63 | 2.9 |
| **Chlortoluron** | 0.9970 | 85 ± 21 | 0.85 | 5.6 | 113 ± 4 | 2.5 | 10 |
| **Clothianidin** | 0.9912 | 83 ± 5 | 14 | 25 | 82 ± 13 | 18 | 30 |
| **Cyanazine** | 0.9990 | 119 ± 6 | 0.70 | 5.5 | 91 ± 10 | 1.6 | 8.8 |
| **DEA** | 0.9990 | 121 ± 8 | 4.0 | 13 | 100 ± 16 | 2.3 | 7.8 |
| **DIA** | 0.9983 | 114 ± 8 | 14 | 50 | 83 ± 10 | 7.3 | 22 |
| **Diazinon** | 0.9995 | 88 ± 10 | 0.17 | 0.58 | 108 ± 8 | 0.43 | 1.8 |
| **Dichlorvos** | 0.9916 | 100 ± 5 | 28 | 40 | 81 ± 3 | 6.8 | 20 |
| **Diflufenican** | 0.9951 | 80 ± 17 | 14 | 49 | BLOD | 120 | 260 |
| **Dimethoate** | 0.9926 | 99 ± 5 | 14 | 46 | BLOD | 180 | 330 |
| **Diuron** | 0.9949 | 117 ± 3 | 0.11 | 0.39 | 87 ± 17 | 0.57 | 2.5 |
| **Fenitrothion** | 0.9810 | 121 ± 6 | 13 | 44 | BLOD | 170 | 300 |
| **Fenitrothion oxon** | 0.9977 | 89 ± 11 | 1.6 | 5.3 | 85 ± 5 | 6.3 | 18 |
| **Fenthion oxon** | 0.9977 | 100 ± 4 | 0.38 | 1.7 | 99 ± 8 | 3.2 | 4.3 |
| **Fenthion oxon sulfone** | 0.9907 | 111 ± 10 | 6.1 | 20 | 93 ± 11 | 19 | 40 |
| **Fenthion oxon sulfoxide** | 0.9963 | 95 ± 9 | 2.2 | 7.4 | 121 ± 4 | 1.9 | 6.4 |
| **Fenthion sulfone** | 0.9830 | 88 ± 16 | 5.6 | 21 | 98 ± 15 | 17 | 39 |
| **Fenthion sulfoxide** | 0.9978 | 85 ± 13 | 1.4 | 4.7 | 127 ± 5 | 3.9 | 9.6 |
| **Fluroxypyr** | 0.9918 | 122 ± 6 | 13 | 42 | 89 ± 14 | 18 | 59 |
| **Imidacloprid** | 0.9949 | 117 ± 4 | 3.9 | 13 | 109 ± 5 | 4.0 | 10 |
| **Irgarol** | 0.9919 | 114 ± 6 | 0.86 | 2.9 | 97 ± 6 | 1.1 | 6.6 |
| **Isoproturon** | 0.9980 | 101 ± 5 | 1.1 | 1.9 | 80 ± 6 | 1.5 | 7.1 |
| **Linuron** | 0.9946 | 119 ±19 | 3.7 | 15 | 89 ± 8 | 3.2 | 12 |
| **Malaoxon** | 0.9917 | 125 ± 17 | 0.88 | 2.5 | 109 ± 9 | 1.8 | 4.1 |
| **Malathion** | 0.9941 | 102 ± 10 | 13 | 44 | 101 ± 10 | 4.1 | 17 |
| **MCPA** | 0.9968 | 90 ± 12 | 2.3 | 7.6 | 91 ± 14 | 2.5 | 6.2 |
| **Mecoprop** | 0.9914 | 85 ± 17 | 3.3 | 11 | 91 ± 11 | 1.7 | 5.7 |
| **Methiocarb** | 0.9903 | 88 ± 11 | 0.28 | 0.95 | 106 ± 8 | 1.7 | 11 |
| **Metolachlor** | 0.9979 | 113 ± 5 | 0.84 | 2.8 | 119 ± 7 | 0.49 | 1.2 |
| **Molinate** | 0.9945 | 125 ± 3 | 16 | 63 | BLOD | 120 | 280 |
| **Oxadiazon** | 0.9914 | 122 ± 10 | 13 | 29 | BLOD | 130 | 440 |
| **Pendimethalin** | 0.9905 | 80 ± 3 | 11 | 37 | BLOD | 190 | 300 |
| **Propanil** | 0.9945 | 125 ± 17 | 2.1 | 7.2 | 100 ± 11 | 6.7 | 20 |
| **Quinoxyfen** | 0.9973 | 87 ± 18 | 0.66 | 2.6 | 109 ± 5 | 5.0 | 16 |
| **Simazine** | 0.9998 | 86 ± 4 | 2.4 | 6.7 | 102 ± 6 | 5.1 | 18 |
| **Terbuthylazine** | 0.9960 | 92 ± 6 | 0.20 | 0.76 | 83 ± 22 | 0.58 | 1.4 |
| **Terbutryn** | 0.9980 | 106 ± 8 | 0.16 | 0.54 | 122 ± | 0.39 | 1.5 |
| **Thiacloprid** | 0.9994 | 112 ± 14 | 0.52 | 1.8 | 115 ± 4 | 0.30 | 0.79 |
| **Thiamethoxam** | 0.9991 | 87 ± 9 | 25 | 33 | 102 ± 8 | 1.8 | 5.9 |
| **Thifensulfuron methyl** | 0.9888 | 124 ± 19 | 1.3 | 4.4 | 88 ± 9 | 1.0 | 3.1 |
| **Triallate** | 0.9934 | 120 ± 20 | 2.6 | 8.6 | 121 ± 6 | 1.1 | 5.1 |

BLOD: Below limit of detection.

**Table 4.** Recovery and repeatability (RSD, relative standard deviation) obtained from the replicate (n=5) analysis of groundwater and surface water fortified with the target analytes at concentrations levels of 10 and 1000 ng/L.

| Analyte | Groundwater | | Surface water | |
|---|---|---|---|---|
| | Analyte recovery ± RSD (%) | | Analyte recovery ± RSD (%) | |
| | 10 ng/L | 1000 ng/L | 10 ng/L | 1000 ng/L |
| **2,4-D** | 91 ± 15 | 95 ± 9 | 126 ± 7 | 127 ± 4 |
| **Acetamiprid** | 80 ± 19 | 101 ± 14 | 93 ± 5 | 80 ± 5 |
| **Alachlor** | 82 ± 11 | 80 ± 16 | 125 ± 5 | 98 ± 6 |
| **Atrazine** | 118 ± 10 | 119 ± 4 | 89 ± 5 | 94 ± 8 |
| **Azinphos ethyl** | 112 ± 17 | 92 ± 13 | 118 ± 17 | 85 ± 8 |
| **Azinphos methyl** | 114 ± 20 | 112 ± 18 | 113 ± 9 | 100 ± 10 |
| **Azinphos methyl oxon** | 121 ± 3 | 90 ± 5 | BLOD | 83 ± 23 |
| **Bentazone** | 107 ± 15 | 86 ± 16 | 104 ± 5 | 81 ± 8 |
| **Bromoxynil** | 82 ± 6 | 82 ± 18 | 121 ± 4 | 90 ± 5 |
| **Chlorfenvinphos** | 113 ± 17 | 90 ± 4 | 120 ± 14 | 93 ± 6 |
| **Chlorpyrifos** | 116 ± 11 | 89 ± 4 | 105 ± 7 | 122 ± 3 |
| **Chlortoluron** | 123 ± 16 | 88 ± 14 | 114 ± 11 | 87 ± 18 |
| **Clothianidin** | BLOD | 99 ± 7 | BLOD | 90 ± 12 |
| **Cyanazine** | 127 ± 24 | 120 ± 20 | 126 ± 4 | 120 ± 13 |
| **DEA** | 81 ± 6 | 93 ± 4 | 90 ± 6 | 102 ± 8 |
| **DIA** | BLOD | 104 ± 8 | 125 ± 3 | 112 ± 6 |
| **Diazinon** | 106 ± 10 | 104 ± 9 | 125 ± 18 | 106 ± 4 |
| **Dichlorvos** | BLOD | 87 ± 3 | 123 ± 7 | 95 ± 5 |
| **Diflufenican** | BLOD | 84 ± 20 | BLOD | 80 ± 17 |
| **Dimethoate** | BLOD | 84 ± 5 | BLOD | 110 ± 9 |
| **Diuron** | 111 ± 16 | 121 ± 4 | 85 ± 17 | 86 ± 11 |
| **Fenitrothion** | BLOD | 98 ± 3 | BLOD | 81 ± 5 |
| **Fenitrothion oxon** | 82 ± 17 | 88 ± 5 | 118 ± 8 | 91 ± 3 |
| **Fenthion oxon** | 107 ± 4 | 124 ± 20 | 102 ± 8 | 115 ± 6 |
| **Fenthion oxon sulfone** | 83 ± 5 | 106 ± 8 | BLOD | 94 ± 17 |
| **Fenthion oxon sulfoxide** | 99 ± 12 | 110 ± 16 | 98 ± 7 | 89 ± 8 |
| **Fenthion sulfone** | 116 ± 4 | 120 ± 4 | BLOD | 83 ± 11 |
| **Fenthion sulfoxide** | 84 ± 3 | 104 ± 8 | 95 ± 5 | 107 ± 11 |
| **Fluroxypyr** | BLOD | 97 ± 6 | BLOD | 79 ± 20 |
| **Imidacloprid** | 84 ± 4 | 83 ± 4 | 112 ± 15 | 118 ± 5 |
| **Irgarol** | 110 ± 15 | 122 ± 16 | 116 ± 10 | 102 ± 11 |
| **Isoproturon** | 104 ± 5 | 83 ± 5 | 107 ± 7 | 119 ± 13 |
| **Linuron** | 92 ± 4 | 106 ± 12 | 109 ± 6 | 86 ± 20 |
| **Malaoxon** | 120 ± 11 | 124 ± 10 | 125 ± 12 | 121 ± 13 |
| **Malathion** | BLOD | 80 ± 20 | BLOD | 88 ± 5 |
| **MCPA** | 115 ± 19 | 86 ± 13 | 93 ± 20 | 113 ± 13 |
| **Mecoprop** | 99 ± 16 | 86 ± 18 | 105 ± 14 | 106 ± 15 |
| **Methiocarb** | 111 ± 15 | 116 ± 19 | 88 ± 11 | 104 ± 12 |
| **Metolachlor** | 122 ± 10 | 114 ± 13 | 122 ± 19 | 118 ± 4 |
| **Molinate** | BLOD | 81 ± 8 | BLOD | 88 ± 7 |
| **Oxadiazon** | BLOD | 109 ± 20 | BLOD | 104 ± 7 |
| **Pendimethalin** | BLOD | 99 ± 4 | BLOD | 121 ± 4 |
| **Propanil** | 120 ± 4 | 85 ± 17 | 103 ± 19 | 110 ± 9 |
| **Quinoxyfen** | 95 ± 3 | 88 ± 21 | 73 ± 5 | 106 ± 19 |
| **Simazine** | 123 ± 12 | 86 ± 20 | 96 ± 4 | 79 ± 19 |
| **Terbuthylazine** | 115 ± 19 | 111 ± 13 | 111 ± 7 | 83 ± 4 |
| **Terbutryn** | 99 ± 3 | 100 ± 3 | 117 ± 13 | 95 ± 8 |
| **Thiacloprid** | 112 ± 15 | 99 ± 6 | 122 ± 3 | 81 ± 18 |
| **Thiamethoxam** | BLOD | 80 ± 6 | 88 ± 12 | 98 ± 12 |
| **Thifensulfuron methyl** | 108 ± 14 | 84 ± 19 | 115 ± 3 | 91 ± 11 |
| **Triallate** | 75 ± 19 | 113 ± 19 | 106 ± 20 | 87 ± 14 |

BLOD: Below limit of detection

**Table 5.** Recovery and repeatability (RSD, relative standard deviation) obtained from the replicate (n=5) analysis of LC-grade water fortified with the target analytes at concentration levels of 10, 100 and 1000 ng/L, and limits of detection (LOD) and determination (LODet) achieved.

| **Analyte** | **Analyte recovery ± RSD (%)** | | | **Sensitivity** | |
|---|---|---|---|---|---|
| | **10 ng/L** | **100 ng/L** | **1000 ng/L** | **LOD ng/L** | **LODet ng/L** |
| **2,4-D** | 79 ± 14 | 81 ± 12 | 88 ± 5 | 6.1 | 20 |
| **Acetamiprid** | 106 ± 9 | 81 ± 19 | 95 ± 9 | 0.16 | 0.53 |
| **Alachlor** | 93 ± 15 | 105 ± 1 | 97 ± 3 | 1.2 | 3.8 |
| **Atrazine** | 102 ± 5 | 123 ± 14 | 92 ± 2 | 0.14 | 0.88 |
| **Azinphos ethyl** | 113 ± 14 | 85 ± 6 | 98 ± 12 | 0.42 | 1.4 |
| **Azinphos methyl** | 81 ± 6 | 92 ± 13 | 121 ± 13 | 0.38 | 1.3 |
| **Azinphos methyl oxon** | 126 ± 10 | 100 ± 7 | 108 ± 11 | 3.1 | 10 |
| **Bentazone** | 76 ± 20 | 88 ± 20 | 113 ± 10 | 4.3 | 14 |
| **Bromoxynil** | 111 ± 5 | 99 ± 5 | 121 ± 8 | 2.6 | 8.6 |
| **Chlorfenvinphos** | 112 ± 11 | 106 ± 4 | 112 ± 15 | 0.24 | 0.80 |
| **Chlorpyrifos** | 123 ± 18 | 120 ± 10 | 104 ± 18 | 0.44 | 1.5 |
| **Chlortoluron** | 125 ± 20 | 98 ± 5 | 107 ± 14 | 0.13 | 0.42 |
| **Clothianidin** | 113 ± 11 | 100 ± 4 | 80 ± 5 | 2.3 | 7.5 |
| **Cyanazine** | 115 ± 7 | 112 ± 5 | 124 ± 3 | 0.081 | 0.28 |
| **DEA** | 90 ± 6 | 100 ± 26 | 102 ± 8 | 2.3 | 7.9 |
| **DIA** | 105 ± 21 | 120 ± 5 | 116 ± 13 | 4.4 | 15 |
| **Diazinon** | 82 ± 3 | 103 ± 5 | 125 ± 6 | 0.042 | 0.16 |
| **Dichlorvos** | 94 ± 20 | 120 ± 15 | 113 ± 14 | 5.4 | 18 |
| **Diflufenican** | 121 ± 11 | 96 ± 13 | 100 ± 19 | 1.2 | 4.0 |
| **Dimethoate** | 120 ± 9 | 117 ± 17 | 84 ± 19 | 0.76 | 2.6 |
| **Diuron** | 109 ± 4 | 127 ± 12 | 124 ± 3 | 0.13 | 0.43 |
| **Fenitrothion** | 106 ± 12 | 123 ± 5 | 120 ± 17 | 2.6 | 8.8 |
| **Fenitrothion oxon** | 120 ± 4 | 85 ± 8 | 112 ± 6 | 0.79 | 2.6 |
| **Fenthion oxon** | 110 ± 12 | 119 ± 3 | 122 ± 7 | 0.17 | 0.59 |
| **Fenthion oxon sulfone** | 99 ± 13 | 125 ± 4 | 112 ± 20 | 2.8 | 9.4 |
| **Fenthion oxon sulfoxide** | 110 ± 4 | 98 ± 5 | 109 ± 6 | 0.13 | 0.43 |
| **Fenthion sulfone** | 85 ± 20 | 109 ± 15 | 120 ± 6 | 4.2 | 14 |
| **Fenthion sulfoxide** | 89 ± 5 | 93 ± 20 | 106 ± 16 | 0.41 | 1.4 |
| **Fluroxypyr** | BLOD | 103 ± 18 | 102 ± 14 | 29 | 95 |
| **Imidacloprid** | 124 ± 20 | 103 ± 12 | 81 ± 18 | 0.87 | 2.9 |
| **Irgarol** | 89 ± 7 | 121 ± 16 | 87 ± 21 | 0.85 | 2.8 |
| **Isoproturon** | 91 ± 24 | 98 ± 16 | 90 ± 3 | 0.15 | 0.50 |
| **Linuron** | 122 ± 3 | 108 ± 8 | 114 ± 8 | 0.58 | 1.9 |
| **Malaoxon** | 90 ± 19 | 117 ± 16 | 123 ± 14 | 0.15 | 0.50 |
| **Malathion** | 83 ± 5 | 82 ± 12 | 82 ± 12 | 3.4 | 12 |
| **MCPA** | 118 ± 6 | 101 ± 20 | 82 ± 7 | 5.5 | 19 |
| **Mecoprop** | 92 ± 17 | 109 ± 4 | 81 ± 13 | 1.1 | 3.6 |
| **Methiocarb** | 110 ± 16 | 123 ± 11 | 108 ± 12 | 0.41 | 1.4 |
| **Metolachlor** | 112 ± 3 | 108 ± 7 | 114 ± 4 | 0.086 | 0.32 |
| **Molinate** | 93 ± 16 | 82 ± 17 | 122 ± 8 | 1.1 | 3.6 |
| **Oxadiazon** | 100 ± 4 | 82 ± 19 | 88 ± 19 | 1.3 | 4.5 |
| **Pendimethalin** | BLOD | 94 ± 4 | 93 ± 7 | 17 | 55 |
| **Propanil** | 122 ± 15 | 112 ± 10 | 112 ± 17 | 0.90 | 3.0 |
| **Quinoxyfen** | 109 ± 12 | 86 ± 3 | 92 ± 7 | 1.1 | 3.6 |
| **Simazine** | 113 ± 12 | 96 ± 20 | 104 ± 13 | 0.31 | 1.1 |
| **Terbuthylazine** | 122 ± 13 | 117 ± 6 | 115 ± 9 | 0.14 | 0.48 |
| **Terbutryn** | 92 ± 11 | 106 ± 4 | 120 ± 4 | 0.19 | 0.66 |
| **Thiacloprid** | 97 ± 3 | 110 ± 18 | 80 ± 14 | 0.059 | 0.21 |
| **Thiamethoxam** | 119 ± 22 | 120 ± 20 | 89 ± 1 | 1.8 | 6.0 |
| **Thifensulfuron methyl** | 83 ± 1 | 118 ± 12 | 80 ± 19 | 0.022 | 0.06 |
| **Triallate** | 114 ± 20 | 110 ± 14 | 118 ± 4 | 3.8 | 13 |

BLOD: Below limit of detection

Analyte recoveries observed in each of the investigated matrices were in general in good agreement at the three concentration levels (Tables 3-5). Analyte losses during extraction and variations in analyte ionization due to matrix effects were well compensated with the use of SIL standards, as indicated by the recoveries obtained, always between 80% and 120%, except in a few cases that slightly deviated from this range. Likewise, relative standard deviations (RSD) nearly always below 20%, or very close, indicated good repeatability, as corresponds to automated methodologies with minimal sample manipulation.

The average LODs and LODets obtained in surface water were between 0.3 and 19 ng/L and between 0.8 and 40 ng/L, respectively, for most of the compounds (86%), while in groundwater these limits ranged between 0.1 and 28 ng/L and from 0.4 to 63 ng/L for all compounds, respectively.

The extent of matrix effects in both surface and groundwater is shown in Figure 4. Significant matrix effects (±20% variation of the signal) were observed for 90% of the compounds in both matrices. MS signal enhancement was observed only for bentazone (+54%) in groundwater and for fenitrothion (+94%), DIA (+37%), and MCPA (+37%) in surface water. In all other cases, matrix effects occurred in the form of signal ionization suppression, with values up to -100%.

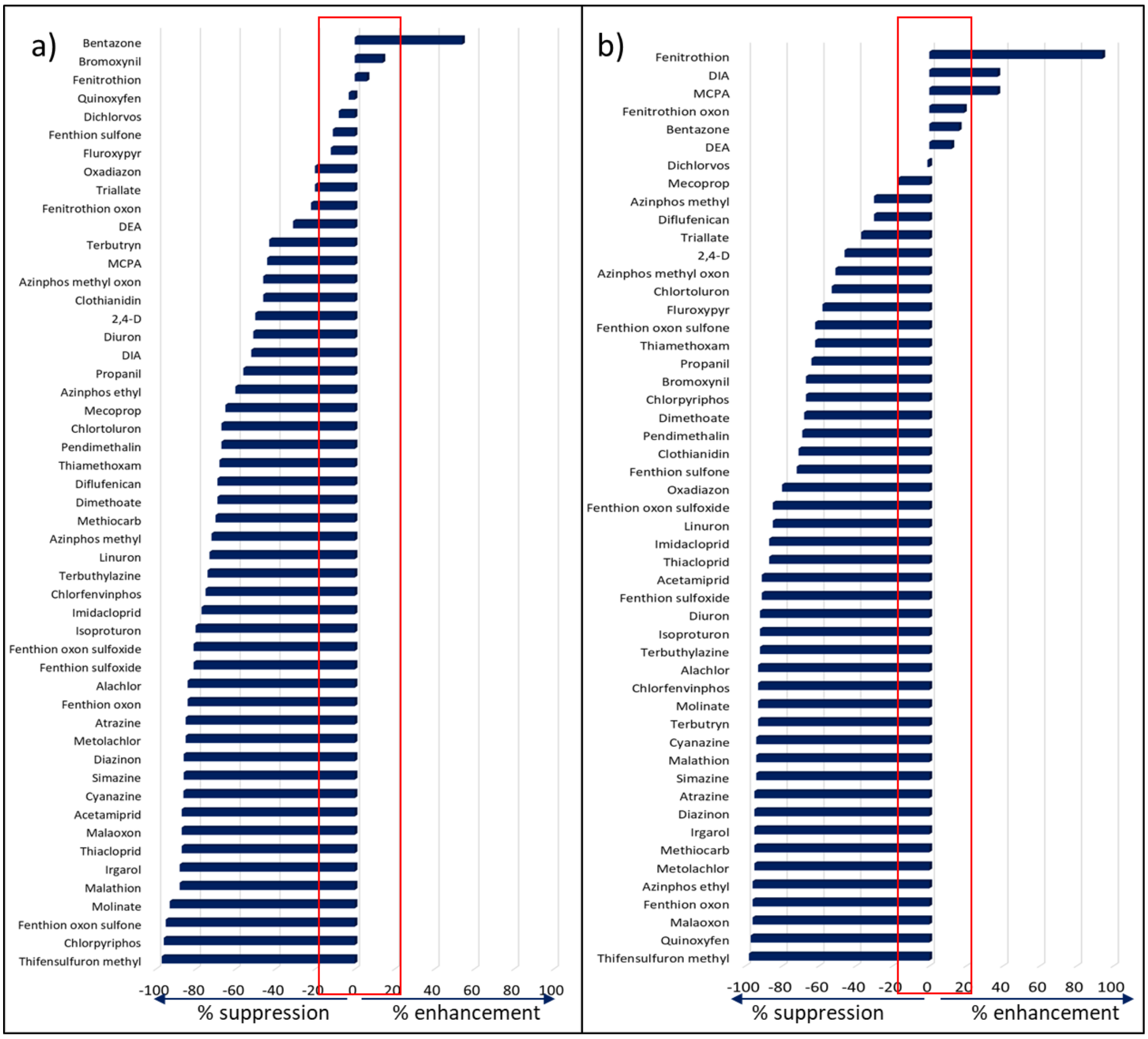

**Figure 4.** Matrix effects observed at a concentration level of 100 ng/L in groundwater (a) and surface water (b).

In the last ten years, of all the methodologies published in the peer-reviewed literature for the determination of non-polar and polar pesticides in water samples, only a few of them are fully automated (Camilleri et al., 2015; Hurtado-Sánchez et al., 2013; Mann et al., 2016; Quintana et al., 2019; Rubirola et al., 2017; Singer et al., 2010) (Table 6). Among them, the present method and the recent one described by Quintana et al. (2019) are the only capable of determining more than 50 pesticides in water samples, covering a wide spectrum of medium to highly polar compounds. A special feature of the methodology here presented as compared to the others is the large proportion of SIL analogs used for quantification (88%). The use of SIL compounds for almost all targeted analytes indeed requires an initial investment of economic resources for their acquisition; however, it is essential to correct for matrix effects and ensure the production of reliable results in any water matrix.

**Table 6.** On-line SPE LC-MS/MS methodologies published in the peer-reviewed literature in the last ten years for the simultaneous determination of medium to highly polar pesticides in groundwater and surface water**.**

| Number of pesticides | Analyte overlap | Water matrix | Pre-extraction step | Quantification method | Accuracy (analyte recovery, %) | Precision (RSD, %) | Sensitivity (LOQ, ng/L) | Matrix effects | Reference |
|---|---|---|---|---|---|---|---|---|---|
| **51** | | GW/SW | centrifugation | isotope dilution (88% of SIL) | . 80-127 | <20 | GW: 0.4-63 SW: 0.8-40 (86% of analytes) | ± 100 | This study |
| **96** | 30 | SW/GW /DW | centrifugation | isotope dilution (22% of SIL) | -40 to 42 | <40 | 5-25 | -161 to 100 | (Quintana et al., 2019) |
| **14** | 13 | SW/DW /EWW | filtration | isotope dilution (86% of SIL) | <10 relative bias[a] | <10 | SW: 0.3 – 2.1 | ± 100 | (Rubirola et al., 2017) |
| **23** | 8 | DW/SW /GW | filtration | isotope dilution (22% of SIL) | 72-198 | <40 | GW: 8-62 SW: 10-64 | not provided | (Mann et al., 2016) |
| **10** | 4 | SW | acidification (2.5‰ formic acid) | not provided | 86-114 | <30 | 0.1-10 | ± 20 | (Camilleri et al., 2015) |
| **37** | 10 | SW | filtration | Standard addition | 74-129 | <14 | 0.3-33 | not provided | (Hurtado-Sánchez et al., 2013) |
| **20** | 9 | SW/WW | filtration | Isotope dilution (60% of SIL) | 71-103 | <20 | SW: 3-100 | ± 100 | (Singer et al., 2010) |

GW, groundwater; SW, surface water; SIL, stable isotope-labeled analogs; DW, drinking water; EWW, effluent wastewater; WW, wastewater.

[a] Relative bias (%) = ((theoretical concentration − experimental concentration)/theoretical concentration) × 100

Overall, on-line methods allow lowering the LODs to a higher extent than off-line analytical approaches because the complete sample (5 mL in our case) is transferred into the LC-MS systems. In comparison with other automated analytical methods available in the literature, the LODets obtained with the methodology developed, between 0.4 and 63 ng/L in groundwater and between 0.8 and 40 ng/L in surface water (for 86% of the compounds), are overall comparable or lower than those previously reported by other authors, for instance: LOQs from 0.3 to 33 ng/L reported by Hurtado-Sánchez et al. for 10 pesticides in surface water (Hurtado-Sánchez et al., 2013), from 3 to 100 ng/L reported by Singer et al. for 20 pesticides also in surface water (Singer et al., 2010), and LOQs values above 8 ng/L in groundwater and 10 ng/L in surface water as reported by Mann et al. (Mann et al., 2016). In this respect, it may be worth mentioning that the LODets in our method incorporate the confirmation by the SRM2, and thus could be higher than the LOQ of the SRM1 if the LOD of the SRM2 is above that value. Despite this, our automated approach provides the best sensitivity for the analysis of azinphos-methyl, bromoxynil, clothianidin, quinoxyfen, terbuthylazine, terbutryn, and thifensulfuron methyl, and to the best of our knowledge, this is the first time that an on-line SPE-LC MS/MS-based method is validated for the analysis of azinphos ethyl, azinpho- methyl oxon, dichlorvos, diflufenican, fenthion oxon, fenthion oxon sulfone, fenthion oxon sulfoxide, fenthion sulfone, fenthion sulfoxide, and oxadiazon in surface and groundwater samples. Moreover, the sensitivity of the presented methodology allows its application to monitor the target priority substances in surface waters below their respective lowest EQS (EC, 2013): alachlor, LOD=6 ng/L *vs.* EQS=300 ng/L; atrazine, 2 ng/L *vs.* 600 ng/L; chlorfenvinphos, 0.6 ng/L *vs.* 100 ng/L; chlorpyrifos, 0.6 ng/L *vs.* 30 ng/L; diuron, 0.6 ng/L *vs.* 200 ng/L; irgarol, 1 ng/L *vs.* 2.5 ng/L; isoproturon, 2 ng/L *vs.* 300 ng/L; quinoxyfen, 5 ng/L *vs.* 15 ng/L; simazine, 5 ng/L *vs.* 1000 ng/L; and terbutryn, 0.4 ng/L *vs.* 6.5 ng/L. Dichlorvos (LOD 6.8 ng/L) is the only target priority substance that cannot be detected with the proposed methodology below its lowest

EQS (0.06 ng/L). The developed methodology also provides LODs in compliance with the maximum acceptable LODs established in the European Watch List (EC, 2018) for the detection of methiocarb (1.7 ng/L *vs.* 2 ng/L) and four of the five neonicotinoids included in our study (imidacloprid 4 ng/L, thiacloprid 0.3 ng/L, acetamiprid 1 ng/L, thiamethoxam 1.8 ng/L, and as an exception clothianidin 18 ng/L, in all cases *vs* 8.3 ng/L). Regarding groundwater, all the studied pesticides can be detected at levels below 0.1 µg/L, which is the quality standard for individual pesticides established in the European Directive for the protection of groundwater against pollution and deterioration (EC, 2006a).

It is also worthy to highlight that, contrary to our method that uses centrifugation for sample pre-treatment, almost all the automated methods evaluated filtrate the water to remove suspended particles before analysis. Centrifugation is as effective as filtration to remove suspended solids; however, it avoids the potential retention of the less polar compounds onto the filters and reduces time and analysis costs.

Besides increased sensitivity and repeatability, this methodology presents as additional advantages over other analytical methods available in the literature for the analysis of polar pesticides: i) full automation (which results in minimum sample preparation and manipulation requirements, i.e., only 10 min centrifugation and SIL standards addition, ii) the use of a low sample volume (5 mL, which simplifies sample transport and storage), iii) high sample throughput (SPE of a sample is conducted during LC-MS/MS analysis of the previous sample in a batch, and the whole process takes only 40 min), iv) cost efficiency (due to low solvents consumption and avoidance of evaporation steps), and v) overall time saving (due to low maintenance and easy operation of the instrument, and automated data processing (MassLynx)).

### *3.3 Occurrence in water samples*

The developed methodology was applied to the analysis of 22 water samples collected in two agriculture-impacted areas of Catalonia with different predominant crops and pressures. The results obtained (lowest and highest concentrations, average concentrations, and detection frequencies) are summarized in Table 7, while individual concentrations of the pesticides found in the investigated samples are provided as SI in Tables 8 and 9.

Of the 51 investigated compounds, 28 were detected in the Llobregat basin. The pesticide pattern observed in the studied area (Figures 2 and 5), which includes small tributaries (Pt 1 and Pt 2) and their confluence into the main river (Pt 3), as well as irrigation and drainage channels that give service to surrounding farms (Pt 4-11) and in some points, receive the input of WWTP effluents (Pt 6 and Pt 9), was characterized by the generalized presence of diuron and terbutryn throughout the investigated stretch, with punctually high concentrations of other compounds in certain sites, such as bromoxynil in Pt 4 and linuron in Pt 7 and 10. A variety of compounds were present in some locations, *viz.*, Pt 7 and Pt 10, in line with the variety of small exploitations dedicated to the cultivation of different crops (Figure 5). In this profile, it was also notable the presence of 2,4-D in the two sites most directly affected by the input of WWTP effluents (Pt 6 and Pt 9), which reflects a likely poor removal of this compound in the WWTPs.

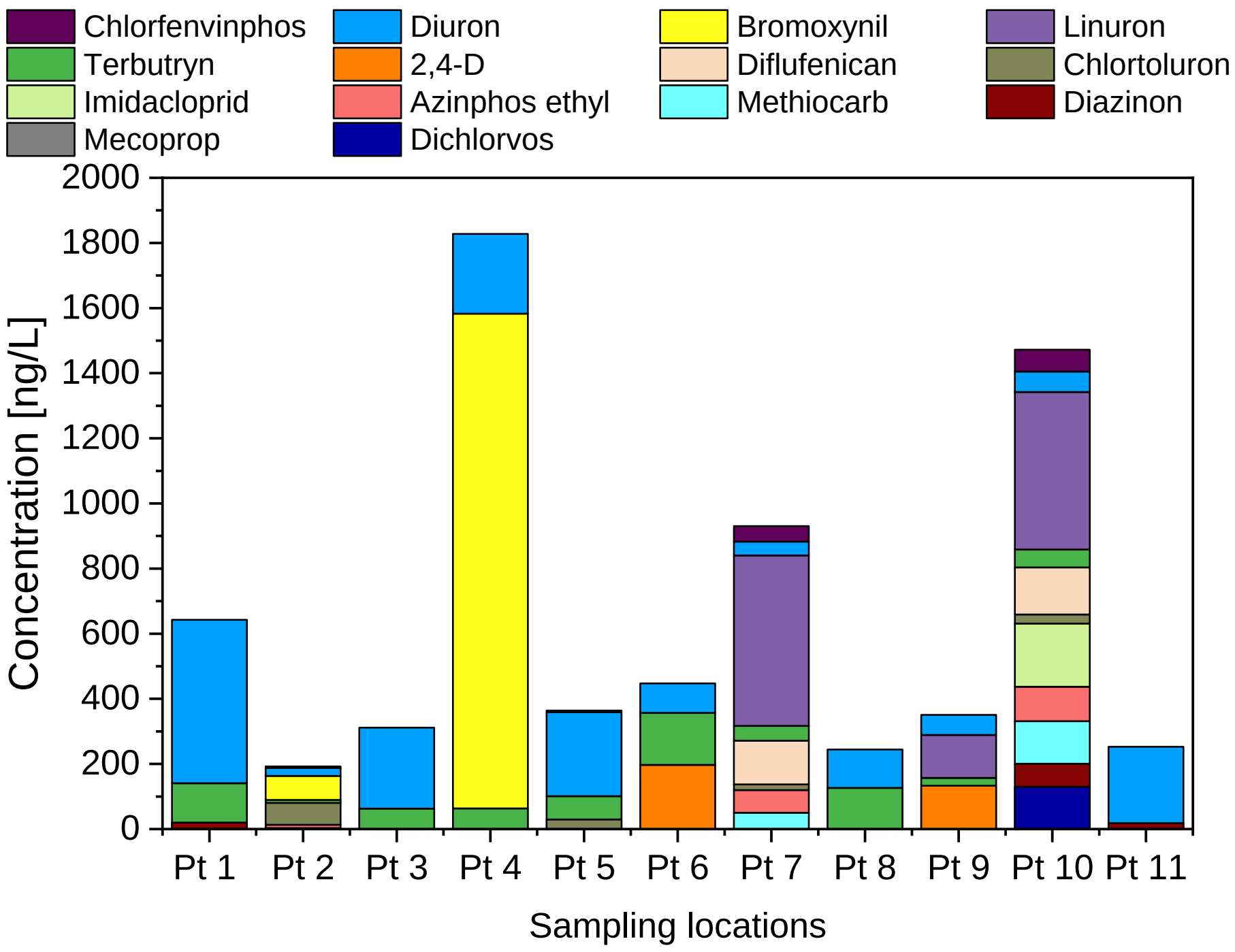

**Figure 5.** Cumulative levels of the most abundant (>100 ng/L in at least one sample) and/or frequently detected (>36%) pesticides found in the lower basin of the Llobregat River. Alachlor, atrazine, cyanazine, fenthion sulfoxide, fenthion oxon, irgarol, isoproturon, malaoxon, malathion, metolachlor, molinate, propanyl, simazine, terbuthylazine, and thiacloprid, detected at lower concentrations in fewer samples, are not represented in the figure.

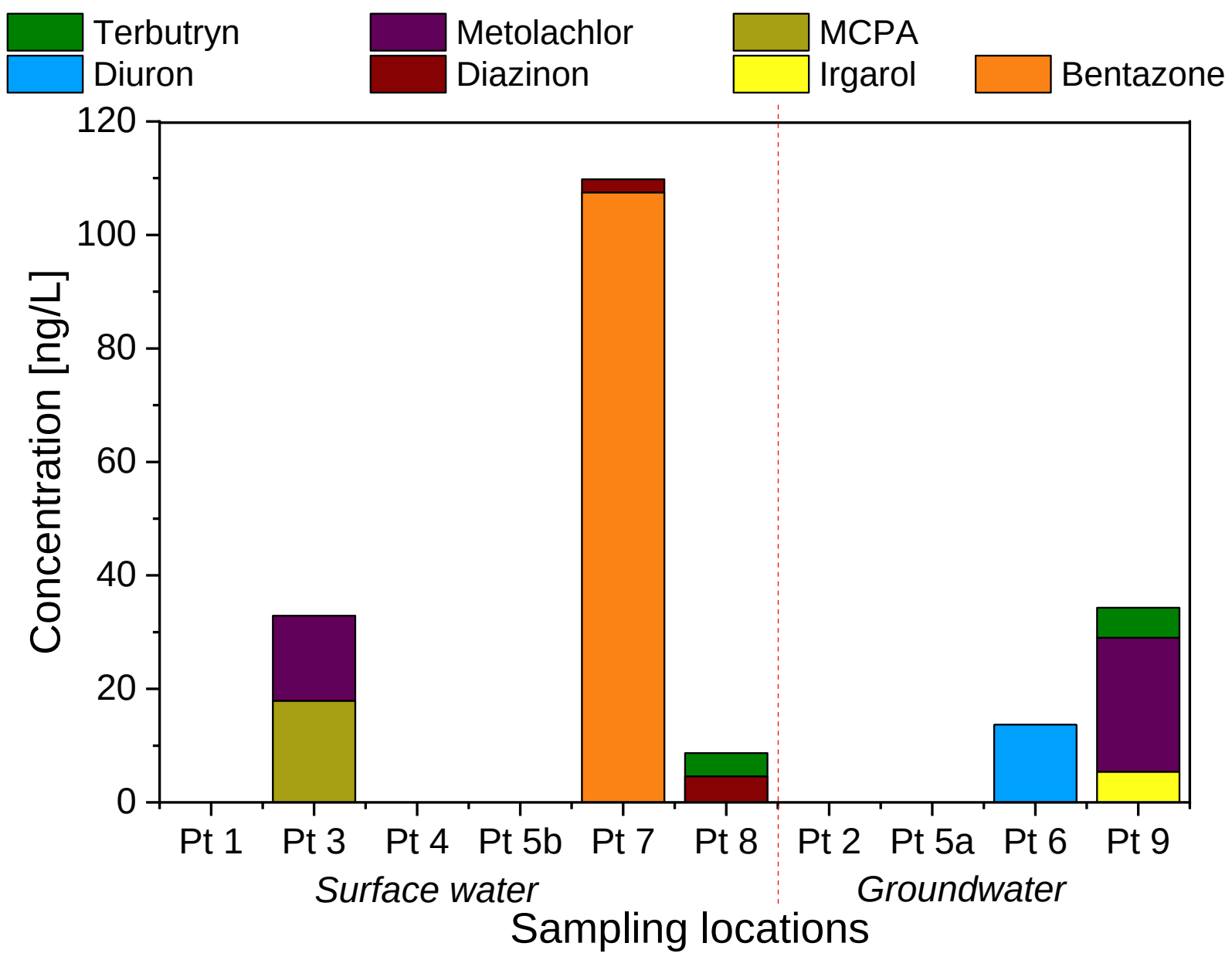

**Figure 6.** Cumulative levels of the targeted pesticides measured in the Ter River water samples.

**Table 7.** Minimum, and mean concentrations and detection frequency of the targeted pesticides in the investigated water samples.

| Pesticides | Concentration (ng/L) | | | Detection frequency[b] (%) |
|---|---|---|---|---|
| | Min | Max | Mean[a] | |
| ***Llobregat River*** | | | | |
| **2,4-D** | 130 | 200 | 30 | 18 |
| **Alachlor** | 18 | 24 | 3.8 | 18 |
| **Atrazine** | <6.7 | 21 | 5.6 | 55 |
| **Azinphos ethyl** | 10 | 110 | 17 | 27 |
| **Bromoxynil** | <22 | 1520 | 150 | 27 |
| **Chlorfenvinphos** | <2.9 | 67 | 12 | 55 |
| **Chlortoluron** | 18 | 67 | 13 | 36 |
| **Cyanazine** | 29 | 29 | 2.6 | 9 |
| **Diazinon** | 2.9 | 71 | 11 | 36 |
| **Dichlorvos** | <20 | 130 | 16 | 45 |
| **Diflufenican** | 130 | 150 | 25 | 18 |
| **Diuron** | 24 | 500 | 170 | 100 |
| **Fenthion oxon** | 37 | 37 | 3.4 | 9 |
| **Fenthion sulfoxide** | 32 | 32 | 2.9 | 9 |
| **Imidacloprid** | <10 | 190 | 19 | 27 |
| **Irgarol** | <6.6 | 41 | 7.6 | 45 |
| **Isoproturon** | <7.1 | 25 | 4.8 | 27 |
| **Linuron** | 130 | 520 | 100 | 27 |
| **Malaoxon** | 24 | 24 | 4.4 | 18 |
| **Malathion** | <17 | 32 | 6 | 27 |
| **Methiocarb** | 50 | 130 | 17 | 18 |
| **Metolachlor** | 5.1 | 28 | 5.3 | 27 |
| **Molinate** | 27 | 33 | 5.5 | 18 |
| **Propanil** | 19 | 19 | 1.7 | 9 |
| **Simazine** | 16 | 20 | 3.3 | 18 |
| **Terbuthylazine** | 3.5 | 30 | 5.2 | 27 |
| **Terbutryn** | 8.9 | 160 | 67 | 91 |
| **Thiacloprid** | <0.79 | 31 | 4.3 | 27 |
| ***Ter River*** | | | | |
| **Bentazone** | 110 | 110 | 9.8 | 9 |
| **Diazinon** | 2.3 | 4.6 | 0.63 | 18 |
| **Diuron** | 14 | 14 | 1.3 | 9 |
| **Irgarol** | 5.4 | 5.4 | 0.49 | 9 |
| **MCPA** | 18 | 18 | 1.6 | 9 |
| **Metolachlor** | 15 | 24 | 3.6 | 18 |
| **Terbutryn** | 4.1 | 5.3 | 0.85 | 18 |

[a] Mean calculated considering values <LOQ as LOQ/2 and values <LOD as zero.
[b] % of positive samples (including values >LOD and <LOQ).

**Table 8.** Concentrations (ng/L) of the individual pesticides and cumulative pesticide concentrations (TOTAL) measured in the water samples collected in the Llobregat River.

| PESTICIDES | Pt 1 | Pt 2 | Pt 3 | Pt 4 | Pt 5 | Pt 6 | Pt 7 | Pt 8 | Pt 9 | Pt 10 | Pt 11 |
|---|---|---|---|---|---|---|---|---|---|---|---|
| **2,4-D** | n.d. | n.d. | n.d. | n.d. | n.d. | 197 | n.d. | n.d. | 133 | n.d. | n.d. |
| **Alachlor** | n.d. | n.d. | n.d. | n.d. | n.d. | n.d. | 18 | n.d. | n.d. | 24 | n.d. |
| **Atrazine** | n.d. | 13 | <6.7 | n.d. | n.d. | <6.7 | 17 | n.d. | n.d. | 21 | <6.7 |
| **Azinphos ethyl** | n.d. | 10 | n.d. | n.d. | n.d. | n.d. | 69 | n.d. | n.d. | 106 | n.d. |
| **Bromoxynil** | n.d. | 74 | <22 | 1520 | n.d. | n.d. | n.d. | n.d. | n.d. | n.d. | n.d. |
| **Chlorfenvinphos** | n.d. | 4.8 | n.d. | n.d. | 4.8 | <2.9 | 48 | n.d. | n.d. | 67 | <2.9 |
| **Chlortoluron** | n.d. | 67 | n.d. | n.d. | 30 | n.d. | 18 | n.d. | n.d. | 27 | n.d. |
| **Cyanazine** | n.d. | n.d. | n.d. | n.d. | n.d. | n.d. | n.d. | n.d. | n.d. | 29 | n.d. |
| **Diazinon** | 20 | 2.9 | n.d. | n.d. | n.d. | n.d. | n.d. | n.d. | n.d. | 71 | 18 |
| **Dichlorvos** | n.d. | <20 | <20 | n.d. | <20 | n.d. | n.d. | <20 | n.d. | 130 | n.d. |
| **Diflufenican** | n.d. | n.d. | n.d. | n.d. | n.d. | n.d. | 134 | n.d. | n.d. | 145 | n.d. |
| **Diuron** | 502 | 24 | 249 | 245 | 258 | 91 | 42 | 118 | 61 | 63 | 235 |
| **Fenthion oxon** | n.d. | n.d. | n.d. | n.d. | n.d. | n.d. | 37 | n.d. | n.d. | n.d. | n.d. |
| **Fenthion sulfoxide** | n.d. | n.d. | n.d. | n.d. | n.d. | n.d. | 32 | n.d. | n.d. | n.d. | n.d. |
| **Imidacloprid** | n.d. | n.d. | n.d. | n.d. | n.d. | n.d. | <10 | n.d. | n.d. | 194 | <10 |
| **Irgarol** | n.d. | n.d. | n.d. | <6.6 | <6.6 | n.d. | 33 | <6.6 | n.d. | 41 | n.d. |
| **Isoproturon** | n.d. | n.d. | n.d. | n.d. | n.d. | n.d. | 24 | <7.1 | n.d. | 25 | n.d. |
| **Linuron** | n.d. | n.d. | n.d. | n.d. | n.d. | n.d. | 524 | n.d. | 132 | 484 | n.d. |
| **Malaoxon** | n.d. | n.d. | n.d. | n.d. | n.d. | n.d. | 24 | n.d. | n.d. | 24 | n.d. |
| **Malathion** | n.d. | n.d. | n.d. | <17 | n.d. | n.d. | 25 | n.d. | n.d. | 32 | n.d. |
| **Methiocarb** | n.d. | n.d. | n.d. | n.d. | n.d. | n.d. | 50 | n.d. | n.d. | 131 | n.d. |
| **Metolachlor** | 5.1 | n.d. | n.d. | n.d. | n.d. | n.d. | 25 | n.d. | n.d. | 28 | n.d. |
| **Molinate** | n.d. | n.d. | n.d. | n.d. | n.d. | n.d. | 33 | n.d. | n.d. | 27 | n.d. |
| **Propanil** | n.d. | n.d. | n.d. | n.d. | n.d. | n.d. | 19 | n.d. | n.d. | n.d. | n.d. |
| **Simazine** | 16 | n.d. | n.d. | n.d. | n.d. | n.d. | n.d. | n.d. | n.d. | 20 | n.d. |
| **Terbuthylazine** | n.d. | n.d. | n.d. | n.d. | n.d. | n.d. | 24 | n.d. | 3.5 | 30 | n.d. |
| **Terbutryn** | 121 | 8.9 | 63 | 63 | 71 | 160 | 45 | 127 | 24 | 55 | n.d. |
| **Thiacloprid** | <0.79 | n.d. | n.d. | n.d. | n.d. | n.d. | 16 | n.d. | n.d. | 31 | n.d. |
| ***TOTAL*** | 664 | 205 | 311 | 1830 | 364 | 448 | 1260 | 244 | 354 | 1800 | 252 |

n.d.: not detected

<LOQ: below limit of quantification

Total concentration calculated considering only values >LOQ.

**Table 9.** Concentrations (ng/L) of the individual pesticides and cumulative pesticide concentrations (TOTAL) measured in the water samples collected in the Ter River.

| PESTICIDES | Pt 1 | Pt 2 | Pt 3 | Pt 4 | Pt 5a | Pt 5b | Pt 6 | Pt 7 | Pt 8 | Pt 9 | Pt 10 |
|---|---|---|---|---|---|---|---|---|---|---|---|
| **Bentazone** | n.d. | n.d. | n.d. | n.d. | n.d. | n.d. | n.d. | 108 | n.d. | n.d. | n.d. |
| **Diazinon** | n.d. | n.d. | n.d. | n.d. | n.d. | n.d. | n.d. | 2.3 | 4.6 | n.d. | n.d. |
| **Diuron** | n.d. | n.d. | n.d. | n.d. | n.d. | n.d. | 14 | n.d. | n.d. | n.d. | n.d. |
| **Irgarol** | n.d. | n.d. | n.d. | n.d. | n.d. | n.d. | n.d. | n.d. | n.d. | 5.4 | n.d. |
| **MCPA** | n.d. | n.d. | 18 | n.d. | n.d. | n.d. | n.d. | n.d. | n.d. | n.d. | n.d. |
| **Metolachlor** | n.d. | n.d. | 15 | n.d. | n.d. | n.d. | n.d. | n.d. | n.d. | 24 | n.d. |
| **Terbutryn** | n.d. | n.d. | n.d. | n.d. | n.d. | n.d. | n.d. | n.d. | 4.1 | 5.3 | n.d. |
| ***TOTAL*** | n.d. | n.d. | 33 | n.d. | n.d. | n.d. | 14 | 110 | 8.7 | 34 | n.d. |

n.d.: not detected

Total concentration calculated considering only values >LOQ

As shown in Table 7, the herbicides diuron and terbutryn were the most ubiquitous pesticides, occurring in 100% and 91% of the samples analyzed, respectively. Diuron is an effective herbicide used to treat invasive vegetation on both agricultural and non-agricultural sites. It is also useful in removing mildew and killing algae. Thus, such a widespread occurrence may be associated with its use in agriculture but also in industrial and urban environments. The ubiquitous presence of diuron in this river has been already reported in previous studies (Köck-Schulmeyer et al., 2012; Masiá et al., 2015). On the other hand, terbutryn presence may be attributed to its release from the river sediments (Barbieri et al., 2019; Masiá et al., 2015) or nearby soils, where it may be accumulated, because the use of this herbicide/algaecide as a plant protection product has been banned for nearly a decade in the EU (EC, 2002). The sorption of terbutryn onto solid particles during its use in the past is supported by its low water solubility (25 mg/L) and its moderately high octanol-water partition coefficient (log Kow=3.7) (Table 1).

Diuron was also one of the targeted pesticides that presented the highest concentrations (up to 500 ng/L), only surpassed by bromoxynil (1520 ng/L) and linuron (520 ng/L). Bromoxynil and linuron are both herbicides used to control annual broadleaf weeds on crop and non-crop sites.

In addition to diuron, bromoxynil and linuron, various other pesticides, namely, 2,4-D, azinphos-ethyl, dichlorvos, diflufenican, imidacloprid, methiocarb, and terbutryn, were found at concentrations above the limit of 100 ng/L set for individual pesticides in water intended for human consumption (EC, 1998). This is a concern since the Llobregat River water is an important source of drinking water for the city of Barcelona and its metropolitan area. Total pesticide concentrations in the Llobregat River waters ranged between 205 and 1830 ng/L, being above the limit set for total pesticides of 500 ng/L in four out of the eleven investigated locations (Pt 1, Pt 4, Pt 7, and Pt 10) (Figures 2 and 5). In Pt 1 and Pt 4, the exceedance is basically due to the large presence of one specific pesticide (diuron and bromoxynil, respectively), whereas in the other sites (Pt 7 and Pt 10) the exceedance is due to the concurrent presence of a variety of pesticides at low concentrations. Overall, the main groups contributing to total pesticide levels were triazines, organophosphorus, ureas, and neonicotinoids (Figure 5).

Many of the pesticides detected are priority substances in the field of water policy (EC, 2013) or are included in the EU Watch List (EC, 2018). EQS exceedances (EQS provided in Table 1) were only observed in the case of the antifouling agent irgarol in two locations (33 and 41 ng/L *vs* its EQS of 16 ng/L) and the insecticide dichlorvos in five locations (from 20 ng/L to 130 ng/L, far above its EQS of 0.7 ng/L). The use of irgarol and dichlorvos is currently prohibited in the EU (EC, 2006b; EC, 2009; EC, 2016). Both substances were also found in the Llobregat River sediments (Barbieri et al., 2019) which may be the source of these pollutants into the river water. While dichlorvos has not been previously investigated in the Llobregat River waters, irgarol was previously reported to occur in a tributary of the Llobregat River at a maximum concentration of 5 ng/L (Quintana et al., 2019).

Methiocarb and the neonicotinoids imidacloprid and thiacloprid were found at concentrations (up to 130, 190 and 31 ng/L, respectively) much higher than the maximum acceptable method LODs established in the Watch List for the monitoring of these substances (2 ng/L for methiocarb and 8.3 ng/L for the neonicotinoids). Given that the acceptable method LODs provided by the implementing decision 2018/840 (EC, 2018) coincide with the substance-specific PNEC in water, the measured concentrations could affect aquatic organisms.

As for TPs, malaoxon was detected in two samples at similar concentration levels than its parent compound, malathion (24 ng/L in the case of malaoxon in both samples *vs* 25 and 32 ng/L of malathion). The presence of malaoxon is of concern, considering that it is 60 times more toxic than its parent compound (Jensen and Whatling, 2010). Moreover, the occurrence of malathion in drinking water sources is also worrisome as it may convert into malaoxon during chlorine-based disinfection of water (Ohno et al., 2008) if it survives to the water treatment train. Two TPs of the currently banned organophosphate insecticide fenthion, namely, fenthion oxon and fenthion sulfoxide, were also detected in one of the sampling locations (Pt.7, Figure 2) at concentrations of 37 ng/L and 32 ng/L, respectively, while the parent compound was not detected in any sample.

In the Ter River, pesticide pollution was much less severe than in the Llobregat River (Figure 6). In total, 7 pesticides (bentazone, diazinon, diuron, irgarol, MCPA, metolachlor, and terbutryn) were detected in this area (Table 7). Total concentrations of pesticides in groundwater were very low (up to 34 ng/L in Pt 9), being slightly higher in surface waters (up to 110 ng/L in Pt 7). Thus, pesticide levels did not exceed in any case the limit of 500 ng/L set in the European legislation for the sum of pesticides in groundwater (EC, 2006a) and waters intended for human consumption (EC, 1998). However, bentazone was found in one of the surface water samples (Pt 7) at a level (108 ng/L) higher than the standard of 100 ng/L set for individual pesticides. Bentazone is extensively used as an herbicide in agriculture and especially in rice fields, and likely to reach water bodies due to its high mobility in soils or via runoff (high water solubility = 7112 mg/L and low log $K_{ow}$ = 0.46). The location where bentazone was found corresponded indeed with a drainage channel of a rice cultivation area, and was the most polluted among the investigated sites.

Regarding the occurrence of priority pesticides in the Ter surface waters, only three were found (terbutryn, irgarol, and diuron), but none of them at concentrations above its corresponding MAC (Tables 1 and 9).

Four compounds currently banned in Europe were also detected, including terbutryn, irgarol, metolachlor, and diazinon (EC, 2002; EC, 2006b; EC, 2016). Their presence may be due to improper use of the stock of these compounds or rather to their release by leaching or runoff from soils or sediments, where the contaminants could have accumulated over time.

Pesticide contamination in the Ter River has been scarcely investigated before. A study conducted in 2001 in the same area revealed the presence of atrazine, DEA, and metolachlor at levels below 100 ng/L in samples from the Ter River after water treatment in a plant (Quintana et al., 2001). The use of these compounds is currently banned in Europe and this could explain the absence of atrazine and its metabolite desethyl atrazine (DEA) in our study, and the low levels of metolachlor (15 ng/L) found in surface water.

### *3.4 Environmental risk assessment*

Hazard quotients (HQs) calculated for the various individual pesticides detected in the samples based on their maximum concentrations measured are provided in Table 10. In the case of the Llobregat River basin, six compounds, namely irgarol, dichlorvos, methiocarb, azinphos ethyl, imidacloprid, and diflufenican presented HQ values above 10, in Pt 10, and thus, they represent a potentially high risk for the aquatic organisms. This risk is associated with their very low PNEC values (≤ 0.01 µg/L) and relatively high concentrations measured in this water sample, which is also one of the most contaminated sites due to the co-occurrence of many pesticides. A moderate risk (1<HQ<10) was estimated for eight other pesticides: 2,4-D, bromoxynil, diazinon, diuron, linuron, malathion, terbutryn, thiacloprid. In the case of bromoxynil, diuron, and linuron, the risk is due to the elevated concentrations sporadically found (>500 ng/L).

In the Ter River basin, the only two compounds that exhibited a potential risk, although low, were bentazone (HQ= 1.08), the pesticide found at the maximum concentration, and irgarol (HQ=1.54), which presents a very low PNEC (0.0035 µg/L).

Note that the above risk assessment considers the worst-case scenario (the highest pesticide concentration measured and the lowest concentration at which effects are not observed) and evaluates individual compounds. If the mixture of the pesticides present in each sample is considered by adding the corresponding HQs calculated based on the corresponding concentrations measured, through the so-called additive model, the sites Pt 2, Pt 6, Pt 7 and Pt 10 of the Llobregat River would be under high risk for the aquatic ecosystems (HQ>10) (see Figure 7). In the case of Pt 7 and Pt 10, the high risk is due to the presence of numerous compounds, with a greater contribution of azynphos ethyl, diflufenican and dichlorvos, while Pt 2 and Pt 6 presented high risk as a consequence of the presence mainly of azynphos ethyl and 2,4-D, respectively. These sites, particularly taking into account that contaminant mixtures may be more toxic than expected based on the sum of the toxicity of the single chemicals present (Hayes, 2019), deserve additional monitoring. On the other hand, no high risk was found in the Ter River samples (Figure 7). Only moderate risk was calculated for Pt 7 and Pt 9, due to the presence of bentazone and irgarol, respectively.

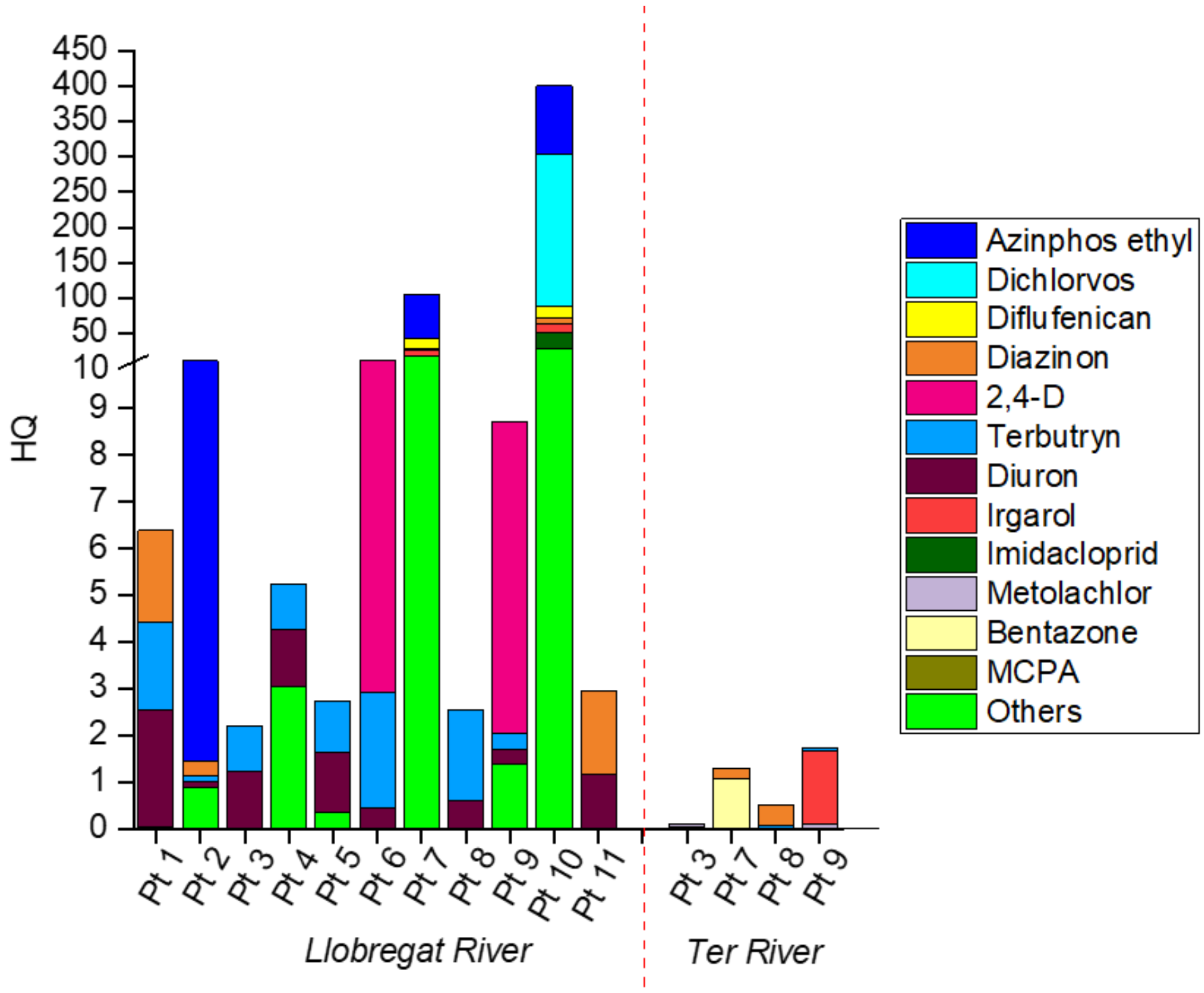

**Figure 7.** Hazard Quotients (HQ) in the samples analysed, based on the individual HQs of the pesticides measured in each sample. The HQs corresponding to those pesticides detected in the samples but not specified in the legend have been grouped as "*Others*".

**Table 10.** Hazard quotient (HQ) values calculated for the pesticides measured in water samples of the Llobregat and Ter River basins.

| Pesticides | MEC[a] (µg/L) | PNEC[b] (µg/L) | HQ |
|---|---|---|---|
| ***Llobregat River*** | | | |
| 2,4-D | 0.20 | 0.02 | 9.86 |
| Alachlor | 0.024 | 0.3 | 0.08 |
| Atrazine | 0.021 | 0.6 | 0.03 |
| Azinphos ethyl | 0.11 | 0.0011 | **96.34** |
| Bromoxynil | 1.5 | 0.5 | 3.04 |
| CFP | 0.067 | 0.1 | 0.67 |
| Chlortoluron | 0.067 | 0.1 | 0.67 |
| Cyanazine | 0.029 | 0.19 | 0.15 |
| Diazinon | 0.071 | 0.01 | 7.08 |
| Dichlorvos | 0.13 | 0.0006 | **216.30** |
| Diflufenican | 0.15 | 0.009 | **16.12** |
| Diuron | 0.51 | 0.2 | 2.51 |
| Fenthion oxon | 0.037 | 0.2 | 0.18 |
| Fenthion sulfoxide | 0.032 | - | - |
| Imidacloprid | 0.19 | 0.0083 | **23.40** |
| Irgarol | 0.041 | 0.0035 | **11.64** |
| Isoproturon | 0.025 | 0.3 | 0.08 |
| Linuron | 0.52 | 0.1 | 5.24 |
| Malaoxon | 0.024 | 0.31 | 0.08 |
| Malathion | 0.032 | 0.006 | 5.30 |
| Methiocarb | 0.13 | 0.01 | **13.06** |
| Metolachlor | 0.028 | 0.2 | 0.14 |
| Molinate | 0.033 | 3.8 | 0.01 |
| Propanil | 0.019 | 0.2 | 0.09 |
| Simazine | 0.020 | 1 | 0.02 |
| Terbuthylazine | 0.030 | 0.06 | 0.51 |
| Terbutryn | 0.16 | 0.065 | 2.45 |
| Thiacloprid | 0.031 | 0.01 | 3.10 |
| ***Ter River*** | | | |
| Bentazone | 0.11 | 0.1 | 1.08 |
| Diazinon | 0.0046 | 0.01 | 0.46 |
| Diuron | 0.014 | 0.2 | 0.07 |
| Irgarol | 0.0054 | 0.0035 | 1.54 |
| MCPA | 0.018 | 0.5 | 0.04 |
| Metolachlor | 0.024 | 0.2 | 0.12 |
| Terbutryn | 0.0053 | 0.065 | 0.08 |

[a] MEC: maximum environmental concentration measured

[b] PNEC: predicted no-effect concentration. Values extracted from [https://www.norman-network.com/nds/ecotox/].

## 4. Conclusions

A fast and simple analytical methodology based on on-line SPE-LC-ESI-MS/MS has been developed for the analysis of a wide range of medium to highly polar pesticides in surface water and groundwater. Advanced aspects of the proposed method are its capability to determine in a single run a high number of multi-class pesticides (51) using a low sample volume (5 mL), in a considerably short analysis time (40 min) and with very high reliability of results (due to the use of isotopically labeled analogs of 45 out of the 51 target compounds for quantification by the isotope dilution method). For most of the initially targeted compounds, the method shows satisfactory performance in terms of accuracy and repeatability and provides enough sensitivity for their detection in ground and surface water in compliance with the current legislation.

The application of the method to real samples showed a very different contamination profile in the two investigated river basins. The average total pesticide concentration in the Ter River samples was 17.6 times lower than in the Llobregat River samples, and only 7 pesticides were found in the Ter River versus 28 detected in the Llobregat River. The list of pesticides found included priority and Watch List substances, and even pesticides currently banned in Europe. The contamination pattern observed in the Llobregat River underlines the significant contribution of the urban and industrial activities conducted in the metropolitan area of Barcelona to pesticide pollution.

High risk for aquatic organisms was expected to be derived from the co-occurrence of many pesticides in specific locations, where pesticides at high concentrations or with very low PNEC values were present. These findings reveal that although less persistent than organochlorine pesticides, medium to highly polar pesticides can be found in water at potentially harmful levels. Further research is needed to understand the sources of these compounds to control them, as well as to assess the real impact of pesticide co-occurrence on the health of the aquatic ecosystems.

**Acknowledgments**

This work has received funding from the European Union's Horizon 2020 Research and Innovation Programme (WaterProtect project, No. 727450), the Spanish State Research Agency (AEI) and the European Regional Development Fund (ERDF) (BECAS project, CTM2016-75587-C2-2-R), the Spanish Ministry of Science and Innovation (Project CEX2018-000794-S), and the Government of Catalonia (2017 SGR 01404). Vinyet Solà from CUADLL and Francesc Camps Saguè from IRTA are acknowledged for their help in the collection of samples from the Llobregat River and the Ter River, respectively.